\documentclass[11pt]{iopart}
\usepackage[colorlinks,linkcolor=blue,citecolor=blue,urlcolor=blue ]{hyperref}

\expandafter\let\csname equation*\endcsname=\relax
\expandafter\let\csname endequation*\endcsname=\relax
\usepackage{amsmath}
\usepackage{graphicx}
\usepackage{enumerate}
\usepackage{color}
\usepackage{bbold}
\usepackage{tensor}
\usepackage{scalerel,stackengine}
\usepackage{empheq}

\def\real{\mathbb R}
\def\id{{\rm 1\kern -2.5pt I}}

\newcommand{\bea}{\begin{eqnarray}} 
\newcommand{\eea}{\end{eqnarray}} 
\newcommand{\dd}{{\mathrm{d}}}

\newcommand{\al}{\alpha} 

\newcommand{\ga}{\gamma}
\newcommand{\De}{\Delta}
\newcommand{\de}{\delta}

\newcommand{\la}{\lambda}

\newcommand{\si}{\sigma}

\newcommand{\vph}{\varphi}
\newcommand{\vep}{\varepsilon}
\newcommand{\ra}{\rightarrow}

\newcommand{\pp}{\partial}

\newcommand{\jacD}[0]{\mathcal{\boldsymbol{D}}}
\newcommand{\jacR}[0]{\boldsymbol{R}}
\newcommand{\sacS}[0]{\boldsymbol{S}}
\newcommand{\EO}[0]{E_{\textrm{o}}}
\newcommand{\ES}[0]{E_{\textrm{s}}}
\newcommand{\laO}[0]{\la_\textrm{o}}
\newcommand{\laS}[0]{\la_\textrm{s}}

\newcommand{\nS}[0]{\boldsymbol{n}_\textrm{s}}
\newcommand{\uO}[0]{\boldsymbol{u}_{\textrm{o}}}
\newcommand{\uS}[0]{\boldsymbol{u}_{\textrm{s}}}

\newcommand{\chiS}[0]{\boldsymbol{\xi}_\textrm{s}}
\newcommand{\thetaO}[0]{\boldsymbol{\theta}_\textrm{o}}

\newcommand{\AS}[0]{A_\textrm{s}}
\newcommand{\rO}[0]{r_\textrm{o}}
\newcommand{\rS}[0]{r_\textrm{s}}
\newcommand{\alphaO}[0]{\alpha_\textrm{o}}
\newcommand{\alphaS}[0]{\alpha_\textrm{s}}
\newcommand{\varepS}[0]{\varepsilon_\textrm{s}}
\newcommand{\varepO}[0]{\varepsilon_\textrm{o}}

\newcommand{\DS}[0]{D_{\textrm{s}}}
\newcommand{\DL}[0]{D_{\ell}}
\newcommand{\DLS}[0]{D_{\ell \textrm{s}}}
\newcommand{\zS}[0]{z_{\textrm{s}}}
\newcommand{\zL}[0]{z_{\ell}}

\newcommand{\sigep}[0]{\sigma_\varepsilon}

\newcommand{\bolde}[0]{\boldsymbol{e}}
\newcommand{\poubelle}[1]{}
\newcommand{\be}{\begin{equation}}
\newcommand{\ee}{\end{equation}}

\definecolor{orange}{rgb}{1,0.5,0}

\newcommand{\giulia}[1]{\textcolor{cyan}{Giulia: #1}}

\begin{document}

\title[Image Rotation from Lensing]{Image Rotation from Lensing }

\author{J\'er\'emie Francfort \& Giulia Cusin \& Ruth Durrer}

\address{D\'epartement de Physique Th\'eorique and Center for Astroparticle Physics,\\
Universit\'e de Gen\`eve, 24 Quai Ernest  Ansermet, 1211 Gen\`eve 4, Suisse}
\ead{jeremie.francfort@unige.ch\\giulia.cusin@unige.ch\\ruth.durrer@unige.ch}

\begin{abstract}
Forthcoming radio surveys will include full polarisation information, which can be potentially useful for weak lensing observations.
We propose a new method to measure the (integrated) gravitational field between a source and the observer, by looking at the angle between the morphology of a radio galaxy and the orientation of the polarisation. For this we use the fact that, while the polarisation of a photon is parallel transported along the photon geodesic, the infinitesimal shape of the source, e.g. its principal axis in the case of an ellipse, is Lie transported as described by the lens map. While at second order, the lens map usually contains a rotation, here we show that the presence of shear alone already induces an apparent rotation of the shape of an elliptical galaxy. As an example, we calculate the rotation of the  shape vector  with respect to the polarisation direction which is generated by a distribution of foreground Schwarzschild lenses. For radio galaxies, the intrinsic morphological orientation of a source and its polarised emission are correlated. It follows that observing both the polarisation and the morphological orientation provides information on both the unlensed source orientation and on the gravitational potential along the line of sight. 
\end{abstract}

\section{Introduction}\label{sec:intro}

All of the forthcoming radio surveys plan to include full polarisation information, which can be potentially useful for lensing observations.
In this article, we discuss the use of the polarisation direction as a proxy for the intrinsic structural direction of a radio galaxy. Indeed, if there exists a correlation (or anticorrelation) between the intrinsic morphological orientation of a source and its polarised emission, then the observed polarisation provides information on the unlensed source orientation.  Such a relationship certainly exists for quasars where the polarisation is closely aligned with the radio jets and this effect has already been exploited to measure gravitational lensing using polarisation observations of quasars, see for example \cite{kronberg1991technique}. The number counts of future deep radio surveys will be dominated by quiescent star-forming galaxies rather than active galactic nuclei. Observations in the local Universe indicate that the orientation of the polarisation of radio emission from these galaxies is orthogonal to the major axis of its ellipticity (see, e.g., \cite{Stil:2008ew,huff:2019cs}). 

The class of sources we have in mind in this analysis are low frequency radio galaxies (typically 1-50 GHz as lower frequencies are significantly depolarised by Faraday rotation~\cite{Mahatma_2020}), for which the dominant source of linear polarisation is expected to be synchrotron radiation due to electrons moving in the magnetic field of the galaxy. For these galaxies, it is reasonable to suppose that the polarisation position angle is related to the overall structure of the galaxy. The origin of the aforementioned correlation is the following: the magnetic field in galaxies is dominantly in the galactic plane (the orthogonal component is very small) and tends to be aligned with galaxy morphology, i.e. the semi-major axis of the galaxy (see e.g. \cite{1966ARA&A...4..245G}). Then polarisation from synchrotron radiation is orthogonal to the semi-major axis (i.e. it is in galactic plane) and orthogonal to the magnetic field component in the galactic plane. Hence it is normal to the galaxy's major axis.  

In this work we introduce a new method to measure the (integrated) gravitational field between a source and the observer by exploiting this correlation. For this we use the fact that in Riemannian geometry there are basically two different ways to transport vectors along a geodesic from the source to the observer: Lie transport and parallel transport. While the polarisation of a photon is parallel transported along the photon geodesic, the infinitesimal shape of the source, given by its principal axis in the case of an ellipse, is Lie transported, leading to the well known Jacobi map~\cite{DiDio:2019rfy,Perlick:2004tq}. We calculate the rotation of the  shape vector  with respect to the polarisation direction which is generated when the light ray passes by a lens. Here we use the fact that a misalignment of the lensing shear and the principal axes of a sheared ellipse induces a rotation of the image. This rotation is typically much larger than the one induced at second order from the rotation-term in the lens map. The latter has been studied in the past, see e.g.~\cite{Sereno:2004cn,Sereno:2004gf,Thomas:2017ewl,Whittaker:2015dp,Whittaker:2014wl} and will not be discussed here. In this paper we study the  case of weak lensing. We derive an analytical expression for the rotation angle between the image semi-major axis and the polarisation vector. 
We then apply it to a simple Schwarzschild lens.
Assuming a distribution of lenses (foreground galaxies) motivated by numerical N-body simulations, we then go on to compute  the probability for the position of the semi-major axis to be rotated by an angle larger than a fixed value. In a future project we shall extend this basic idea to the stochastic cosmological  potential from gravitational clustering.

A related idea has been recently studied in \cite{Brown:2010rr}, in a cosmological context. There, the authors use polarisation as a proxy to separate better intrinsic alignment and shear from weak lensing. In~\cite{huff:2019cs} the use of polarisation information to reduce 'shape noise' in weak lensing measurements is investigated. In the present work, we study the change of the angle between the orientation of the galaxy and polarisation as an observable on its own right. We propose to use this rotation angle (multiplied by the square of the eccentricity of the ellipse) as a new observable.

The paper is organised as follows: in the next Section we collect textbook results on (weak) lensing  which we use in Section~\ref{sec:obs} to determine the rotation angle, denoted as $\de\al$, of the principal axes of an image (ellipse) with respect to a parallel transported  polarisation vector. In Sections \ref{sec:ss} and \ref{sec:stoch} we focus on the case of a Schwarzschild lens, we derive analytical expressions for the rotation angle $\de\al$  in this case and we compute the optical depth to a given  rotation $\de\al$ seen in a random sample of sources, lensed by foreground galaxies. We discuss our results and we conclude in Section~\ref{sec:con}. For the convenience of the reader, some detailed derivations are collected in several appendices.\\ \\

\noindent
\textbf{Notations and conventions:}\\
We work with the metric signature $(-,+,+,+)$. We use the natural system of units in which $c=\hbar=1$, hence $t$, $r$ and $m\equiv GM$ have dimensions of length, where $G$ and $M$ are respectively Newton's constant and the lens mass. We denote the Schwarzschild radius of a lens by $r_{m} = 2 GM=2m$. The affine parameter $\lambda$ has units of length (or time). The relationship between the aspect ratio $a$ and the eccentricity $\varepsilon$ of an ellipse (describing the galaxy shape) is
\begin{equation} 
\varepsilon = \sqrt{1-\frac{1}{a^2}}\,.
\end{equation} 
For an ellipse with eccentricity at the observer $\varepO$ and a rotation angle $\delta \alpha$, we define the \emph{scaled rotation} $\Theta \equiv \vert \delta \alpha \vert \ \varepO^2$. \\
The indices $A$, $B$ go from $1$ to $2$, parametrising the Screen space. If the indices are omitted, 4-vectors are denoted in bold face.
The principal axes of the shear are denoted as $(\bolde_-,\bolde_+)$. 
Below we collect the various angles which we will introduce in the text. 
\begin{enumerate}
    \item The Jacobi matrix is parametrised by the angles $\phi$ and $\psi$, corresponding respectively to the angle between the Sachs basis and the principal axes of the shear and to the net rotation. Note that $\psi$ vanishes in the Schwarzschild case.
    \item The angle between the semi-major axis of the ellipse and the first Sachs vector of the Sachs plane at the source (respectively observer) position is $\alphaS$ (respectively $\alphaO$).
    \item The rotation angle is $\delta \alpha \equiv \alphaO-\alphaS $.
    \item The two polar angles used in spherical coordinates are $\theta$ and $\varphi$.
\end{enumerate}

\section{Weak lensing of elliptical galaxies: basic notions}
In this Section, we recall some concepts and definitions of the lensing theory and we focus then on the case of weak lensing. More details can be found in many different references, for example \cite{Perlick:2004tq, Schneider:2006,Fleury:2015hgz}. We present the parametrisation of the Jacobi matrix used in this work.

\subsection{Lensing in a nutshell}
We first present the formalism of lensing. 
We consider a photon geodesic with wave vector $k^\mu$. If the beam of photons is polarised at the source in some direction $\nS$, this direction is normal to both, the source velocity vector $\uS$ and the photon geodesic $\boldsymbol{k}_{\textrm{s}}$. Hence $n_\textrm{s}^\mu k_\mu^{\textrm{s}} =
n_\textrm{s}^\mu u^{\textrm{s}}_\mu =0$.\footnote{More precisely, the polarisation vector direction is defined by the electric field, given in terms of the Faraday tensor by $E^{\mu}=F^{\mu\nu}u^{\textrm{s}}_{\nu}$. Hence by construction it is a space-like vector normal to the source 4-velocity $\bf u^{\textrm{s}}$, i.e. $E^{\mu}u^{\textrm{s}}_{\mu}=0$. It also lies in the plane orthogonal to the wave vector, hence it has to be orthogonal to both $k_\mu^{\textrm{s}}$ and $u_\mu^{\textrm{s}}$.} Here and in rest of this article, we denote with a label '$\textrm{s}$' quantities defined at the source position while  the label '$\textrm{o}$' denotes quantities defined at the observer. The photon polarisation is parallel transported along the photon geodesic,
\begin{equation}
\dot{\boldsymbol{n}} = 
\frac{D \boldsymbol{n}}{\dd\la}=
\nabla_{\boldsymbol{k}} \boldsymbol{n} = 0 \,,
\end{equation}
where $\la$ is an affine parameter for the photon geodesic $x^\mu(\la)$ such that $k^\mu(\la)=\dd x^\mu(\la)/\dd \la$.
 
We assume the source image to be elliptical.  The end points of its major axis define the vector $\xi^\al$ which obeys the geodesic deviation equation
\begin{equation}\label{e:geodev1}
\ddot \xi^\al=(\nabla_{\boldsymbol{k}}\nabla_{\boldsymbol{k}}\boldsymbol{\xi})^\al=   -R^\al_{\beta\mu\nu}k^\beta \xi^\mu k^\nu\,.
\end{equation}
In order to define the so called Sachs basis~\cite{sachs:1961zz,Perlick:2004tq,Straumann:2013spu} we introduce the 2-dimensional 'screen' at the observer as the 2D part of tangent space which is normal to the observer 4-velocity $\uO$ and the photon velocity $\boldsymbol{k}_{\textrm{o}}$. We define two basis vectors $\bolde_1$ and $\bolde_2$ in this plane such that 
\begin{equation}
\label{eq:sachsbasisdef}
\bolde_A\cdot \bolde_B =\de_{AB}\, \qquad 
\bolde_A\cdot \boldsymbol{k}(\laO) = \bolde_A\cdot \uO = 0 \,, \qquad A\,, B \in\{1\,,\, 2\}\,.
\end{equation}
We parallel transport this basis along the (time reversed) photon geodesic back to the source. 
The screen vectors are parallel transported only if the 4-velocity $u$ is. If we consider a source/observer velocity field that is not parallel transported along the photon geodesic (as in our application in Section~\ref{sec:ss} where we choose $u\propto \pp_t$), the vectors $\bolde_A$ are not strictly speaking parallel transported, but they satisfy $\nabla_{\boldsymbol{k}}\bolde_A \propto \boldsymbol{k}$, see~\cite{Fleury:2015hgz} for details. This defines the vector fields $\bolde_1(\la)\; \bolde_2(\la)$.
Along the photon geodesic, the 2-dimensional subspace of tangent space
\begin{equation}
S(\la) = \left\{ c_1\, \bolde_1(\la)+c_2\, \bolde_2(\la) |  c_i \in \real \right\}\,,
\end{equation}
is called the 'screen' (or the 'Sachs screen') and the vectors $(\bolde_1,\bolde_2)$ are a Sachs basis of the screen.

At the source position, in general, $\uS$ is not normal to the screen and therefore the emitted polarisation is not necessarily in the screen. Denoting by $\Pi(\la)$ the projection of a vector 
onto the screen, we define the 'observable' polarisation direction of the photon beam by
$\boldsymbol{n}_{\mathrm{obs.}} = \Pi(\laS) \nS$, where $\nS$ is the original photon polarisation direction. In coordinates, introducing $\boldsymbol{u}(\la)$ as the observer 4-velocity parallel transported along the photon geodesic such that $\boldsymbol{u}^\mu(\laO)=\uO^\mu$ we define the projector on the screen\footnote{Note that $\boldsymbol{u}(\laS)$ is, in general not the source velocity $\uS$.}
\begin{equation}
{\Pi^\mu}_\nu(\la) = 
\de^\mu_\nu 
- \frac{k^\mu k_\nu }{(k^\alpha u_\alpha)^2} 
- \frac{ k^\mu  u_\nu + k_\nu  u^\mu}{(k^\alpha u_\alpha)} \,,
\end{equation}
where the $\lambda$ dependence is understood. It is indeed easy to verify that ${\Pi^\mu}_\nu   k_\mu = {\Pi^\mu}_\nu   u_\mu=0$, and ${\Pi^\mu}_\nu{\Pi^\nu}_\al = {\Pi^\mu}_\al$.

Let us consider a geodesic deviation vector in the screen,  $\boldsymbol{\xi}(\la)=\xi^A(\la)\bolde_A(\la)$ with screen components denoted by $\xi^A$.  The relation, which maps a direction $(\theta^1,\theta^2)$ on the observer screen into a geodesic deviation at the position $\la$ along the photon geodesic is called the Jacobi map. Since the geodesic deviation equation given by Eq.~\eqref{e:geodev1} is a linear differential equation, we can define a linear map, the Jacobi map $\boldsymbol{D}$, through 
\begin{equation}
\xi^A(\la) = {D^A}_B(\la)\theta^B \,.
\end{equation}
 The geodesic deviation equation (Eq.~\eqref{e:geodev1}) following differential equations for $\boldsymbol{D}$ (see for example~\cite{Perlick:2004tq,Straumann:2013spu})
\begin{align}
  &  \ddot{\jacD} =  \jacR\jacD \,, \label{jacDdotdot} \\
   & \dot{\jacD} = \sacS  \jacD \label{jacDdot}\,,\\
  &  \dot \sacS + \sacS^2 = \jacR\label{sdot}\,,
\end{align}
where a dot denotes derivation with respect to the affine parameter $\lambda$ and  $\jacR$ is given by contractions of the Ricci and Weyl tensors (respectively $R_{\mu\nu}$ and $C_{\mu\nu\rho\sigma}$) with the photon four vector and Sachs basis,  or explicitly 
\begin{align} \label{eq:RPhiPsi}
\jacR&=\left(\begin{array}{cc}
\Phi_{00}&0\\
0&\Phi_{00}
\end{array}
\right)+
\left(
\begin{array}{cc}
-\operatorname{Re}(\Psi_0)&\mathrm{Im}(\Psi_0)\\
\mathrm{Im}(\Psi_0)&\mathrm{Re}(\Psi_0)
\end{array}
\right)\,,
\end{align}
with
\begin{align}
\Psi_0&=-C_{\mu\nu\rho\sigma}k^{\nu}k^{\rho}\overline{m}^{\mu}\overline{m}^{\sigma}\,, \\
\Phi_{00}&=-\frac{1}{2}R_{\mu\nu}k^{\mu}k^{\nu}\,,
\end{align}
where we defined  $\boldsymbol{m} = (\bolde_1 + \mathrm{i}\bolde_2)/\sqrt{2}$ and an overbar indicates complex conjugation. The notation in \eqref{eq:RPhiPsi} is chosen to agree with the Newman-Penrose formalism, see~\cite{chandra}. The quantity $\Phi_{00}$ generates contraction or expansion, while the $\Psi_0$ generates both, shear and rotation. For more details, see for example  \cite{Perlick:2004tq} or \cite{Fleury:2015hgz}.

\subsection{Jacobi matrix parametrisation} \label{sec:jacobimatrixrep}

We now focus on the case of weak lensing where lensing does not generate multiple images. This applies when deflections are weak, typically for images which are well outside the Einstein radius of a given lens-source configuration.  In this case the Jacobi matrix is invertible and its eigenvalues are positive.
The polar decomposition theorem states that, like any invertible $2\times 2$ matrix, the Jacobi map can be decomposed as a unitary times a symmetric matrix. If both eigenvalues of the symmetric matrix are positive, this decomposition is of the form (see \ref{sec:appmatrix} for more details).
\begin{equation}\label{eq:parJ}
 \jacD= \DS
\left(
\begin{array}{cc}
\cos\psi&-\sin\psi\\
\sin\psi&\cos\psi
\end{array}
\right)
\exp\left(\begin{array}{cc}
-\gamma_1&\gamma_2\\
\gamma_2&\gamma_1
\end{array}\right)\,,
\end{equation}
where 
\begin{equation}
\exp\left(\begin{array}{cc}
-\gamma_1&\gamma_2\\
\gamma_2&\gamma_1
\end{array}\right)=
\left(
\begin{array}{cc}
\cos\phi&-\sin\phi\\
\sin\phi&\cos\phi
\end{array}
\right)
\left(\begin{array}{cc}
e^{-\gamma}&0\\
0&e^{+\gamma}
\end{array}\right)
\left(
\begin{array}{cc}
\cos\phi&\sin\phi\\
-\sin\phi&\cos\phi
\end{array}
\right)\,.
\end{equation}
Again, more detailled explanations can be found in \cite{Perlick:2004tq} and \cite{Fleury:2015hgz}, for example. Note that to first order in $\ga$ we have
\begin{align}
\exp\left(\begin{array}{cc}
-\gamma_1&\gamma_2\\
\gamma_2&\gamma_1
\end{array}\right)
=
\left(\begin{array}{cc}
1-\gamma_1&\gamma_2\\
\gamma_2&1+\gamma_1
\end{array}\right) =\left(
\begin{array}{cc}
1- \ga \cos (2 \phi) & -\ga \sin (2 \phi) \\
- \ga \sin (2 \phi) & 1+\ga \cos (2 \phi) \\
\end{array}\label{e:ga-ga2}
\right) \,,
\end{align} 
where
\begin{equation}
 \ga_1 = \ga \cos (2 \phi) \,, \quad  \ga_2 = \ga \sin (2 \phi) \,.
\end{equation}
In a more compact notation we can write the general Jacobi map as
\begin{equation}
 \jacD = \DS R(-\psi-\phi) \exp(-\gamma \sigma_3) R(\phi)\,,
\label{eq:jacobicompact}
\end{equation}
where $R(\chi)$ denotes the (clockwise) 2D rotation by an angle $\chi$,
\begin{equation}
R(\chi) =
\begin{pmatrix}
\cos \chi &  \sin \chi \\ 
-\sin \chi& \cos \chi 
\end{pmatrix}\, \qquad \mbox{and}\qquad \si_3 = \begin{pmatrix}
1 & 0 \\ 
0& -1
\end{pmatrix}\,,
\end{equation}
is the third Pauli matrix. Combining all those results, we obtain
\begin{equation}
\label{eq:thetatoksi}
\boldsymbol{\xi}(\lambda)=  \DS R(-\psi-\phi) \exp(-\gamma \sigma_3) R(\phi) \thetaO = 
 \jacD \thetaO \,.
\end{equation}
The quantities involved in the above decomposition (\ref{eq:parJ}) can be interpreted in the following way. Starting from an observed image, the intrinsic properties of the source are reconstructed by applying the following steps.
\begin{enumerate}
    \item The rotation given by $\phi$ rotates the Sachs basis (counterclockwise) $({\bolde}_1,{\bolde}_2)$ into the 
    principal axes of the shear, $({\bolde}_-,{\bolde}_+)$.  
    \item Assuming $\gamma>0$, these axes are contracted (respectively expanded) by a factor $e^{-\gamma}$ (respectively $e^{\gamma}$).
    \item The image is globally rotated (counterclockwise) by an angle $\psi$.
    \item The factor ${\DS}=\sqrt{\det \jacD}$ converts angles at the observer into length at position $\boldsymbol{x}_{\text{s}}$. In a cosmological setting, this corresponds to the angular diameter distance to the source. 
\end{enumerate}

\section{Observable quantities describing the image rotation}\label{sec:obs}

In this Section, we discuss the observables  describing the rotation of an image, due to lensing. We show that, in the presence of  polararization information, the angle between the galaxy morphology and polarisation direction at the observer describes the image rotation due to lensing, and we present analytical expressions for this observable. Detailed derivations are presented in \ref{sec:appmaps}.

\subsection{Rotation from shear}

We assume that a given source projected on the Sachs screen has an elliptical shape. We shall call this 'the image at the source'. The first observable we can in principle extract is the eccentricity of the observed ellipse. Moreover, assuming that the source emission is  polarised, one can also measure the angle between the semi-major axis and the direction of polarisation. Assuming, e.g., these two directions to be aligned (or orthogonal) at the source position, a non-vanishing (or non $90^o$) angle is a signature of lensing.

Naively, one might expect that the quantity $\psi$ in the Jacobi matrix Eq.~\eqref{eq:jacobicompact}, represents the  rotation of the photon beam.
This type of rotation has been considered in the past, e.g. in~\cite{Sereno:2004cn,Sereno:2004gf,Thomas:2017ewl,Whittaker:2015dp,Whittaker:2014wl}. For a Schwarzschild lens it vanishes identically.
Furthermore, the rotation described by the angle $\psi$ is a second order effect and is in general irrelevant for weak gravitational fields, see for example~\cite{DiDio:2019rfy,Lepori:2020ifz}.
Here we show that the shear $\gamma$ also induces a rotation of the semi-major axis of the ellipse relative to the Sachs basis  if the principal axes of the shear, parametrised by the angle $\phi$ of the Jacobi matrix, denoted $(\bolde_-, \bolde_+)$, do not coincide with the axes of the ellipse.

\begin{figure}[ht!]
    \begin{center}
        \includegraphics[width=0.5\columnwidth]{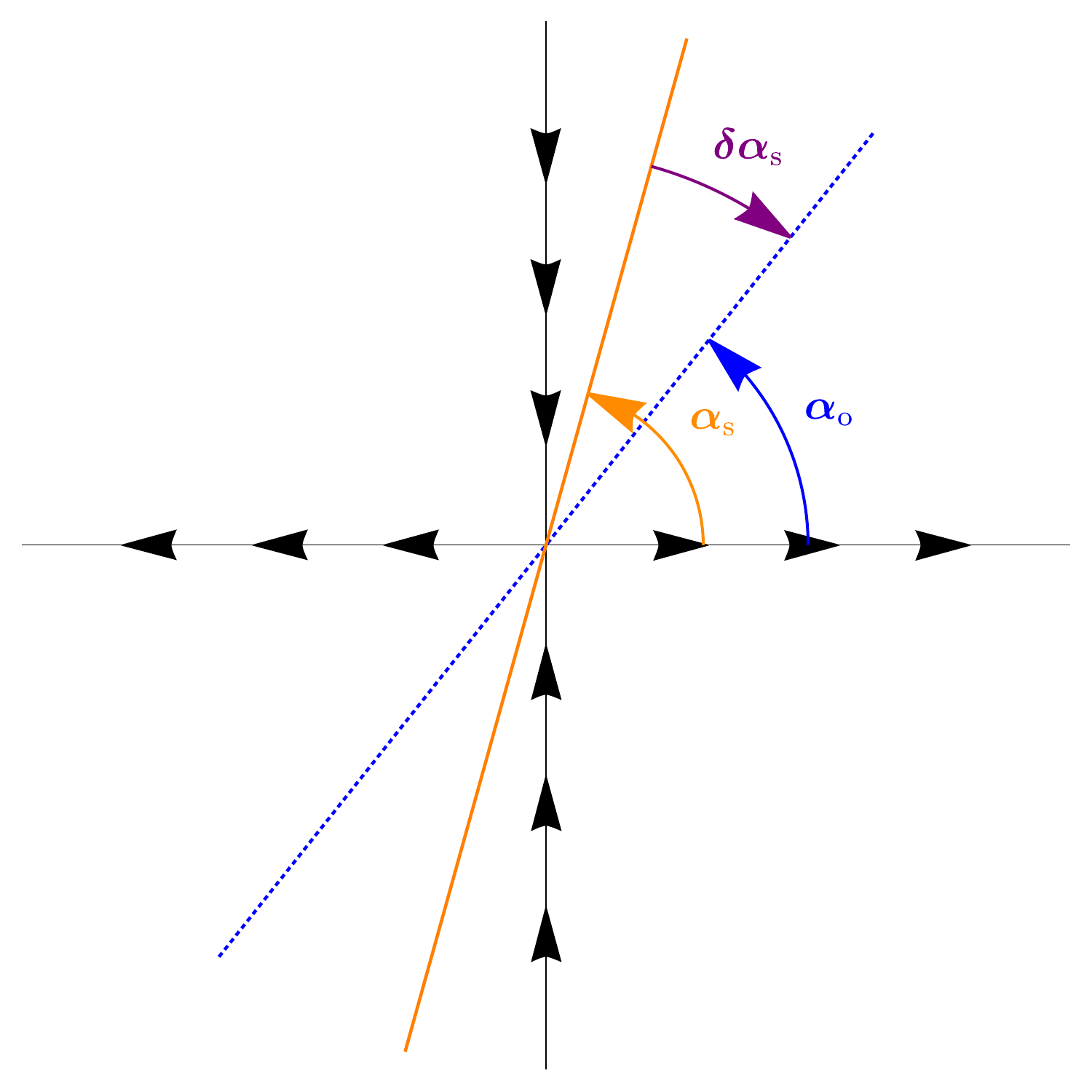}
        \caption{\label{fig:shear_effect} The solid orange line represents the image at the source, supposed to be an ellipse with eccentricity $\varepS\approx 1$. The blue dotted line represents the image at the observer position. They respectively form an angle $\alphaS$ and $\alphaO$ with the horizontal axis. The difference is $\delta \alpha$ and is negative in this example. The arrows on the axes represent the effect on the shear \emph{from} the source image \emph{to} the observed image: The horizontal axis is elongated and the vertical axis is contracted.}
    \end{center}
\end{figure}

This can be intuitively understood with the following simple example, schematically represented in  Fig.~\ref{fig:shear_effect}~:  Consider an ellipse at the source with an eccentricity close to unity $\varepS \approx 1$, such that it can be approximated by a straight line, forming an angle $\alphaS$ with the first Sachs vector. The upper left point of the image position at the source is described by the coordinates $(x_\textrm{s},y_\textrm{s})$ with $y_\textrm s= x_\textrm s \tan \alpha_\textrm s$. Assuming that the source is lensed, and that only the shear $\gamma$ is non-vanishing, using the weak-lensing relation  Eq.~\eqref{eq:thetatoksi}, the upper left point of the image at the observer is given by $(x_\textrm o, y_\textrm o)=(x_\textrm s e^{+\gamma}, y_\textrm s e^{-\gamma}) =(x_\textrm o, x_\textrm o \tan \alpha_\textrm o)$. It is straightforward to solve this for $\alphaO$ to first order in $\ga$. One obtains $\alphaO = \alphaS - \gamma \sin(2\alphaS)\equiv \alphaS + \delta \alpha$ at first order in $\gamma$.  

\subsection{Main results}

We present now the main result of this article. Detailed derivations are given in  \ref{sec:appmaps}. The starting point is an ellipse-shaped source galaxy. The observables are defined on the Screen space at the observer position (e.g. on Earth).\footnote{A precise treatment would consider the projection of the real, intrinsic shape of the galaxy onto the Sachs plane at the source level. However,  we assume here that the alignment properties of the polarisation with the shape of the galaxy hold already at the level of the projected image of the source.}
The signal emitted undergoes lensing described by the Jacobi formalism in terms of the three lensing parameters $\psi$, $\phi$ and $\gamma$ (as defined in Section \ref{sec:jacobimatrixrep}). At the source position we define $\varepS$ as the eccentricity of the image at the source and $\alphaS$ as the angle between the semi-major axis and the first Sachs vector $\bolde_1$. These parameters are different at the observer position due to the lensing. 
More precisely, the eccentricity $\varepO$ and the angle between the semi-major axis and the first Sachs vector $\alphaO$ at the observed position are given by (see \ref{sec:appmaps}) 
\begin{equation}
\label{eq:epsOalphaO}
\varepO = 
\varepS  + \gamma \cos \left[2 (\alphaS-\phi)\right]\frac{2 (1-\varepS^2)
}{\varepS}\,, \quad
\alphaO = \alphaS - \psi - \gamma \sin \left[2 (\alphaS-\phi)\right] \frac{2-\varepS^2}{\varepS^2}\,.
\end{equation}
As mentioned in the previous Section, the angle $\alphaO$ is in general not a physical observable by itself as the orientation of the Sachs basis is arbitrary. The angle $\alphaS$ corresponds to the angle between the semi-major axis and the first Sachs vector at the source position. Assuming that the polarisation of the photon is aligned with the semi-major axis, and using that the polarisation of light is parallel transported along the way, this is as well the angle between the polarisation direction and the first Sachs basis at the observer; it is also not observable by itself. However, the  difference
\begin{equation}\label{eq:StoO-Sec}
\delta \alpha \equiv \alphaO - \alphaS=- \psi - \gamma \sin \left[2 (\alphaS-\phi)\right] \frac{2-\varepS^2}{\varepS^2}\,,
\end{equation}
corresponds to the angle between the semi-major axis of the ellipse and the polarisation direction at the observer position. This is an observable and  a genuine sign of lensing. Note that at first order $\psi$ vanishes, and we may replace $\varepS$ by $\varepO$. Measuring $\de\al$ for many, differently oriented ellipses therefore, in principle allows us to measure $\ga$ and $\phi$.

The reasoning is the same if we consider polarisation along the semi-minor axis. Moreover, the result given in Eq.~\eqref{eq:StoO-Sec} agrees with the heuristic result derived in the previous Section for $\varepS\approx 1$ and $\psi \approx 0$.

\section{An example: The Schwarzschild lens}\label{sec:ss}

We now consider the example where the geometry is given by the Schwarzschild metric, describing a point-like lens. In this case, given the trajectory of a photon, one can define a canonical basis for the Sachs screen. Moreover, the Jacobi equation can be solved exactly at first order in the lens mass, and we can compute analytically the rotation of the source image with respect to the Sachs basis (i.e. the polarisation vector).

\subsection{Setup}

As usual, the Schwarzschild metric is given by 
\begin{equation}
\boldsymbol g = \mathrm{d}s^2 = - \left( 1- \frac{2m}{r} \right)\; \dd t^2+
 \left( 1- \frac{2m}{r} \right)^{-1}\dd r^2 + r^2\; \dd \theta^2 + r^2 \sin^2 \theta\; \dd \varphi^2\,.
\end{equation}
Here $m\equiv GM$, where $M$ is the mass and $G$ is Newton's constant. 

\begin{figure}[ht!]
    \begin{center}
        \includegraphics[width=0.7\columnwidth]{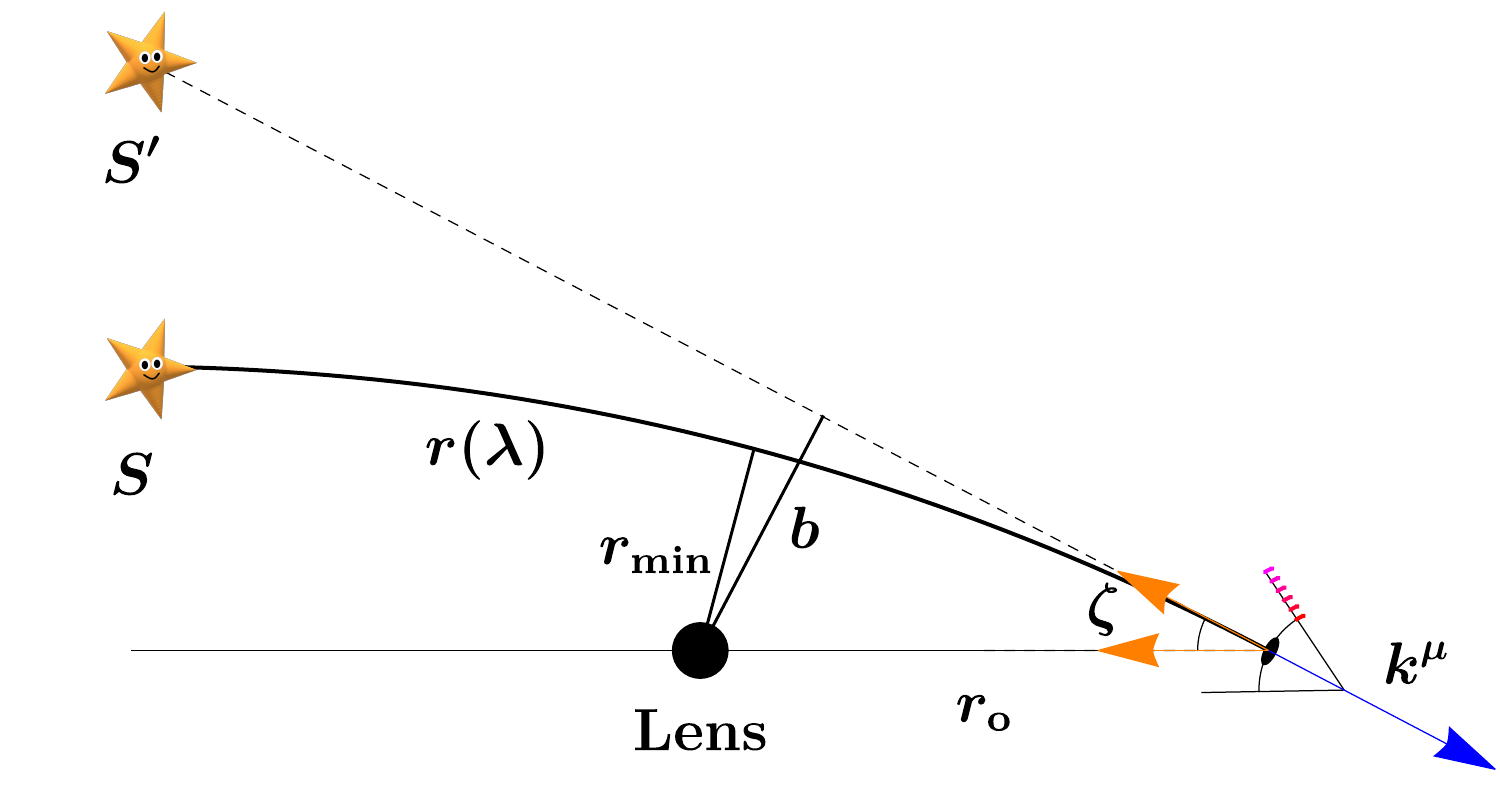}
        \caption{\label{f:ss-setup}A Schwarzschild lens system: the lens corresponds the black circle. The source is located at S and the image is located at S'. The observer is represented by an eye with eyelash. The angle seen by the observer between the lens and the line of sight is $\zeta$. The impact parameter of the trajectory is $r_\textrm{min.}$ while the impact parameter of the unperturbed trajectory is $b$.}
    \end{center}
\end{figure}

We consider the lensing geometry shown in Fig.~\ref{f:ss-setup}. The source is located at an angle $\zeta$ with respect to the optical axis and has impact parameter $b$. The observer is at a distance $\rO$ from the lens. 
To determine the Sachs basis and the Jacobi map, we need the 4-velocity of a set of observers. Since the metric is static, it makes sense to consider the 4-velocity along the timelike Killing vector $\pp_t$,
\begin{equation}
\boldsymbol{u} =\frac{1}{\sqrt{ 1- \frac{2m}{r}}}\pp_t\,.
\end{equation}
Note that the parallel transported vector is not simply  proportional to the time vector $\partial_t$. The induced metric on the spatial surfaces of constant time is $h_{\mu\nu} = g_{\mu \nu} + u_\mu u_\nu$. \\
The 4-momentum of the photon is 
\begin{equation}
\boldsymbol{k} = \frac{E}{1-\frac{2m}{r}}\pp_t +\dot r\pp_r +\frac{L}{r^2}\pp_\varphi\,,
\end{equation}
where $E$ is the dimensionless conserved photon energy and $L$ is its conserved angular momentum (with dimension of a length, as we chose the photon affine parameter to have dimension of  length). One can renormalise the affine parameter to impose $E=1$, which we will assume from now on. The radial equation of motion is given by the condition $\boldsymbol{k}^2 = 0$ and reads
\begin{equation}
\label{eq:rdot}
\dot r = \pm \sqrt{1 - \frac{L^2}{r^2} + \frac{2mL^2}{r^3}} \,,
\end{equation}
where the sign will be different for ingoing and outgoing trajectories. The impact parameter, in the limit $m\ra 0$, is $b=L$. It is possible to express the minimum radius of the perturbed trajectory as a series in $m$. The result is very well-known, see for instance \cite{Renzini:2017bqg}, and at next to leading order is
\begin{equation}
r_\textrm{min.} = b \left( 1 - \frac{m}{b} + \mathcal O\left[\left(\frac{m}{b}\right)^2\right] \right).
\end{equation}
The energies of the photon measured at the source and at the observer respectively are given by
\begin{equation}
\ES = -\boldsymbol{k}\cdot \uS
= \frac{1}{\sqrt{ 1- \frac{2m}{\rS}}}
\,, \qquad 
\EO = -\boldsymbol{k}\cdot \uO
= \frac{1}{\sqrt{ 1- \frac{2m}{\rO}}} \,.
\end{equation}
If both, $\rS\,,\rO\gg 2m$, we have $\ES\simeq \EO\simeq E=1$.

\subsection{The Sachs basis}

We determine now the Sachs basis both, at the observer and at the source positions. Recall that the conditions the Sachs vectors at the observer must satisfy are given by Eq.~\eqref{eq:sachsbasisdef}, and that they are parallel transported along the 4-velocity $\boldsymbol{k}$. For a generic metric, determining the Sachs basis is difficult and may not be possible analytically. However, due to the highly symmetric spacetime considered here, a simple analytical expression exists. For the first Sachs vector, as the geodesic motion lies in the $(r,\varphi)$ plane it seems natural to consider the normalised spatial vector normal to this plane
\begin{equation}
\label{e:s1}
\bolde_1=r^{-1}\pp_\theta\,.
\end{equation}
\begin{figure}[ht!]
    \begin{center}
        \includegraphics[width=0.8\columnwidth]{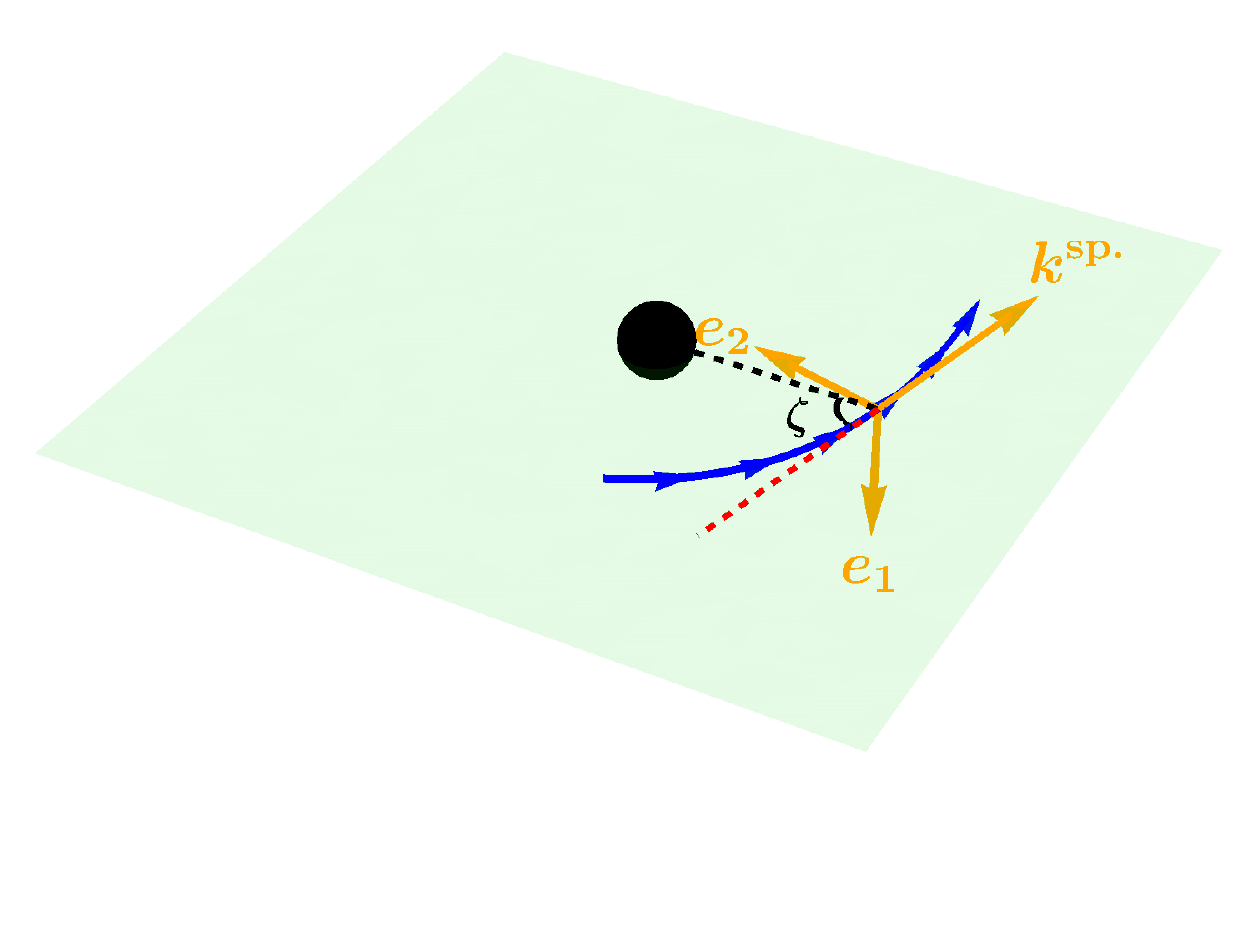}
        \caption{\label{f:ss-sachs}The Sachs basis: The Equatorial plane is in green. The trajectory of the photon is shown in blue and indicated with arrows. The first Sachs vector $\bolde_1$ is normal to the plane while the second Sachs vector $\bolde_2$ is in the plane and has components along $r$ and $\varphi$. It's direct from the figure to see that the system $(\bolde_e, \bolde_2 , \boldsymbol{k}^{\textrm{sp.}})$ is direct (up to their norms).}
    \end{center}
\end{figure} 
Using the definition of parallel transport, it is straightforward to check that $k^\mu \nabla_\mu e_1^\alpha~=~0$, hence this vector is parallel transported. For the second Sachs vector we make the Ansatz
\begin{equation}\label{e:s2}
\bolde_2 = \rho_r \pp_r + \rho_\varphi \pp_\varphi,
\end{equation}
which directly implies $\bolde_1 \cdot \bolde_2 = 0$. Imposing $\bolde_2\cdot \boldsymbol{k}=0$ and $\bolde_2^2=1$ we obtain
\begin{equation}
\rho_r = \frac{b (2 m-r)}{r^2}\,,  \label{e:beta}
\end{equation}
and
\begin{equation}
\rho_\varphi = \pm \frac{1}{r} \sqrt{1- \frac{b^2}{r^2} + \frac{2mb^2}{r^3}} \,,  \label{e:si}
\end{equation}
where the sign has to be the same as the one in the expression for the time derivative of $r$ given by Eq.~\eqref{eq:rdot}. Since the above conditions  do not change under parallel transport and since they determine $\bolde_2$ uniquely, it follows that $\bolde_2$ is parallel transported along $\boldsymbol{k}$ (up to a factor proportional to $\boldsymbol{k}$). This can also be checked by direct calculation. We also introduce the spatial part of the photon momentum
\begin{equation}
k^\textrm{sp.}_\mu \equiv h_{\mu\nu}k^\nu\,,
\end{equation}
where $h_{\mu\nu}$ is the spatial part of the metric
\begin{equation}
h_{\mu \nu} = g_{\mu \nu} + u_\mu u_\nu\,. 
\end{equation}
It is straightforward to check that $(\bolde_1, \bolde_2 , \boldsymbol{\hat k}^{\textrm{sp.}})$ form a direct orthonormal spatial frame, see Fig.~\ref{f:ss-sachs}. More details are given in \ref{sec:apportho}.

\subsection{The Jacobi map}

At the observer position, the angle $\zeta$ represents the angle between the direction of the ingoing photon and the direction to the lens. One finds
\begin{equation}
\label{eq:sinzeta}
\sin\zeta = \frac{b}{\rO} \sqrt{1- \frac{2m}{\rO}} = \frac{b}{\rO} + \mathcal O\left[ \left(  \frac{m}{\rO}\right)\right] \,.
\end{equation}
The origin of the first term is apparent from the geometry of Fig.\,\ref{f:ss-setup}, while the order $m$ correction is derived in \ref{sec:apportho}. 

As we will show in this Section, the Jacobi map is of first order in $m/\rO$. Hence, we only need to determine the trajectory at the unperturbed level (Born approximation), for a given angle $\zeta$. The unperturbed trajectory with $L=b \neq 0$ (and again, using $\EO =1$) is given by
 \bea
 t(\la) &=&\la \,, \\ 
 \varphi(\la) &=& \arccos\left(\frac{b}{\rO}\right)-\arctan\left(\frac{\sqrt{\rO^2-b^2}}{b}+\frac{\lambda}{b}\right) \,, \\
 r(\la) &=& \sqrt{\la^2+\rO^2 +2\la\sqrt{\rO^2-b^2}} \,. \label{eq:rla}
 \eea
 This solution is found imposing that, at the observer, $\la = 0$ and the trajectory with
\begin{equation}
(t,r,\phi_o) =  (0,\rO,0)\,,
\end{equation}
 corresponds to a straight line with impact parameter $r_\textrm{min.} =b$. We consider that $\rO>b$, which implies that the closest approach, $r(\la_{\rm{min.}})=b$ is reached for $\lambda_{\rm{min.}} = - \sqrt{\rO^2- b^2} <0$. We also consider that $\la$ moves forward in time, hence the source is 'located' at $\la<0$. 
Correspondingly, the unperturbed Sachs basis is given by \eqref{e:s1} and \eqref{e:s2} when neglecting $m$ in \eqref{e:beta} and \eqref{e:si}. 
As the Ricci tensor vanishes, the matrix $R$ is given by $\Psi_0$ alone. A short calculation gives
\begin{equation}
\jacR = \Psi_0\left(\begin{array}{cc} -1 & 0 \\ 0 & 1\end{array}\right) \,, \qquad  
\Psi_0 = \frac{3mb^2}{r^5} \,.
\end{equation}
Note that this result is actually exact, as it is obtained making use of the full expression for $\bolde_i$ and $\boldsymbol{k}$. 
With the initial conditions $\jacD(0)=0$ and $\dot\jacD(0)=\mathbb{1}$, Eq.~\eqref{jacDdotdot}  yields
\begin{equation}
\label{eq:fplusminus}
\jacD = \begin{pmatrix}  f_+ &0 \\ 0 &  f_- \end{pmatrix} \qquad \mbox{with} \qquad 
f_\pm = \la \pm m f_\pm^{(1)}\,.
\end{equation}
 To first order in $m$, the differential equation satisfied by the functions $f_\pm$ is 
\begin{equation}
m\ddot f_\pm^{(1)} = \mp\la\Psi_0\,,
\end{equation}
and the initial conditions are
\begin{equation}
f^{(1)}_\pm(0) = \dot f^{(1)}_\pm(0) =0\,.
\end{equation}
From these relations, it is clear that
\begin{equation}
f^{(1)}_- = - f_+^{(1)},
\end{equation}
and it is then sufficient to solve
\begin{equation}
\ddot f_+^{(1)} =-\frac{3b^2 \la}{r^5}, \qquad
f_+^{(1)} = \dot f_+^{(1)} =0.
\end{equation}
We  use flat space expression $r(\la)$ given by Eq.~(\ref{eq:rla}) as the function $f_+^{(1)}$ is already first order in $m/r$. The analytic solution for $f_+^{(1)}$ can be obtained but it is not very illuminating, and we will work in the following in the regime in which the impact parameter $b$ is small with respect to the distance to the lens and to the source, $b/\rO\ll 1$, $b/\rS\ll 1$. 
When the source is behind the lens, as we typically assume, ($\rO+\lambda_{\text{s}} <0$), one has
\begin{equation} \label{eq:fbehind}
m f_+^{(1)}(\la) =-m \frac{4\rO}{b^2}(\rO+\lambda)+\mathcal O(b^2)\,.
\end{equation}
In  the  literature  on  weak  gravitational  lensing  the Jacobi  map  is often  written as $\jacD=\bar{D}_{\textrm{s}} \mathcal{A}$,  
where $\mathcal{A}$ is called the amplification matrix and $\bar{D}_{\textrm{s}}$ is the background angular diameter distance to the source (in our non-expanding case all distances are equivalent and $\bar{D}_{\textrm{s}}=-\lambda_{\textrm{s}}$). Then the magnification  is simply given by
\begin{equation}
\mu=\frac{1}{\det \mathcal{A}}\,.
\end{equation}
Notice that for vacuum solutions ${\bf{R}}$ is traceless and therefore also the Jacobi matrix, traceless at first order, so that magnification is a second order effect. In our case it is also diagonal.

At first order, the Jacobi map (\ref{eq:parJ}) is given by 
\begin{align}
\jacD &= \la_s
\begin{pmatrix}  \exp(-\ga) &0\\ 
0 &  \exp(\ga) \end{pmatrix}\,,
\end{align}
where $\gamma$ is the shear and
\begin{equation}\label{det}
\vert \lambda_s \vert = - \lambda_s= \sqrt{\det \jacD}\,,
\end{equation}
represents the distance to the source, and can be approximated by the angular distance of the source in a cosmological context. We only consider the most interesting case in which the source is behind the lens given by Eq.~\eqref{eq:fbehind}. 
Then making use of (\ref{det}), to lowest order in the impact parameter $\lambda_s = -(\rO+\rS)$ so that the shear is given by 
\begin{equation}\label{shear}
\ga = -  m \frac{f_+^{(1)}(\lambda_s)}{\lambda_s} \simeq \frac{4m\rO}{b^2}\left(\frac{\rS}{\rO+\rS}\right) \,.
\end{equation}

\subsection{Validity of the result}

At first we might be somewhat surprised by the divergent behavior of \eqref{shear} for $b\ra 0$. But of course the weak lensing regime, within which this discussion is valid, requires that the angular position of the source with respect to the optical axis, $\zeta$, is larger than the Einstein angle, i.e. 
\begin{equation}\label{zeta}
\zeta \gg \theta_{\textrm{E}}\,, \qquad \theta_{\textrm{E}} = R_{\textrm{E}}/\rO = \left(\frac{4m\rS}{\rO(\rO+\rS)}\right)^{1/2} \,.
\end{equation}
In this regime one of the two images of the Schwarzschild lens is strongly demagnified. In a more realistic situation with a non-singular lens so that the odd image theorem applies~\cite{Schneider:2006}, Eq.~\eqref{zeta} corresponds to the regime where there is only one image. Here $R_E$ is the Einstein radius and $\theta_E$ is the Einstein angle of the configuration. 
Comparing Eq.\,\eqref{shear} with \eqref{zeta} and making use of Eq.~\eqref{eq:sinzeta}, we obtain
\begin{equation}
\ga \sim  \left(\frac{\theta_E}{\zeta}\right)^{2}  \ll 1\,.
\end{equation}
in the weak lensing regime.

\subsection{Rotation of the beam}
We now consider a source with eccentricity $\varepS$ and whose minor-axes forms an angle $\alphaS$ with the first Sachs vector. The rotation angle becomes to lowest order, see Eq.~\eqref{eq:StoO-Sec}, with $\psi=0$
\begin{equation}
\label{master}
\delta \alpha =  \frac{4m \rO}{b^2}\left(\frac{\rS}{\rO+\rS}\right) \sin\left(2(\alphaS-\phi) \right) \frac{2-\varepS^2}{\varepS^2}\,.
 \end{equation}
If $\alphaS=\phi$ the source ellipse is aligned with the Sachs basis and there is no rotation.
While $\al$ is not well defined when $\varepS=0$, it is nevertheless interesting to note that the rotation is largest for small eccentricity. Intuitively, this means that when the eccentricity is small, for fixed shear $\gamma$ we have to rotate the ellipse by a larger angle to find its new principal axes.

To obtain typical numbers we consider the light of a galaxy passing another galaxy (the lens) of mass $M\sim10^{11}M_\odot$ positioned at $\rO=10\; \mathrm{Mpc}$ away from us with an impact parameter $b=10\;\mathrm{kpc}$ and source located at $\rS\gg \rO$. With this we can write
\begin{equation}
\ga \simeq 1.9\times10^{-3}\left(\frac{M}{10^{11}M_\odot}\right)\left(\frac{r_o}{10{\rm Mpc}}\right)\left(\frac{10{\rm kpc}}{b}\right)^2 \,.
\end{equation}
For an eccentricity of $\vep\sim 1/2$ and angles $\alphaS-\phi$ with $\sin(2(\alphaS-\phi))\sim 1/2$ this corresponds to a rotation angle $\delta \al\sim 0.38^o$ which is not so small. For significantly smaller eccentricity $\delta \al$ or  a large lensing galaxy with $M\simeq 10^{12}M_\odot$ can also be much larger.

\section{The stochastic signal}\label{sec:stoch}
We now explore this rotation in a cosmological context, and we promote all distances to angular diameter distances. In particular the distance observer-lens $\rO\rightarrow \DL$ and the distance observer-source $r_s=-\lambda_s \rightarrow  \DS$. 
In~\ref{ap:error} we show that a fixed error in  measuring the size of the semi-major axis $a$ leads to an error $\propto \varepsilon^{-2}$ in its orientation. For this reason we define a new observable, called the \emph{scaled rotation} 
\begin{equation}
\Theta = \vert \delta \alpha \vert\  \varepO^2 \simeq  
\vert \delta \alpha \vert \ \varepS^2\,,
\end{equation}
which we can measure with a well controlled error.
With these modifications, and orienting our Sachs basis such that  $\phi=0$, the expression for this observable becomes, using eq.\,(\ref{master})
\begin{align}
\Theta  &= \frac{4m\DL}{b^2} \frac{\DLS}{\DS} F(\alphaS, \varepS)\quad \textrm{with}\\
F(\alphaS, \varepS) &= \vert \sin(2 \alphaS)\vert ( 2-\varepS^2) \,.
\end{align}
The impact parameter of the source (in the lens plane) is given by $|{\boldsymbol{b}}|=b=\zeta \DL$. A source with impact parameter equal to $b$ acquires a rotation angle $\delta \alpha$ given by Eq.\,(\ref{master}). 
Hence the cross section for a scaled rotation larger than $\Theta$  from a point like lens is 
\begin{equation}\label{e:cross}
\sigma(\Theta, \varepS, \zL, \zS, m)
= \pi b^2 
=  \frac{4 \pi m \DL}{ \Theta }  \left(\frac{\DLS}{\DS}\right) F(\alphaS, \varepS)\,.
\end{equation}
Using the redshift to parametrise the path between the observer and a source at redshift $\zS$, the optical depth can be expressed as
\begin{equation}\label{tau}
\tau( \Theta , \zS) = 
\int\, p(x)n^{\text{phys}}(z,M)\frac{\mathrm{d}r}{\mathrm{d}z} \sigma( \Theta , x, z,\zS,M)\; \mathrm{d}M \;\mathrm{d}^2x \; \mathrm{d}z\,,
\end{equation}
where $x=(\alphaS, \varepS)$ denotes the intrinsic parameters of the source and we keep the corresponding probability distribution function (PDF) $p(x)$ unspecified for the moment. The physical number density of lenses with mass $M$ at redshift $z$ is given by $n^{\text{phys}}(z,M)$. Recall that $m=GM$ has dimension of a length. The integral over redshift in eq.\,(\ref{tau}) runs from $z=0$ to $\zS$, i.e. it takes into account the contributions of lenses located between the observer and the source. 
The integral over the intrinsic parameters can be factorised out, and the optical depth reads  
\begin{align}
\label{eq:tau}
\tau( \Theta , \zS) &= \mathcal{A} \int\,  \frac{1}{(1+z)H(z)}
\left( \frac{4\pi GM \DL}{ \Theta } \right)
\frac{\DLS}{\DS}
n^{\text{phys}}(z,M)\; \mathrm{d}z\;\mathrm{d}M\,,
\end{align}
where we used that the derivative of physical distance with respect to redshift is given by 
\begin{equation}
\frac{\mathrm{d}r}{\mathrm{d}z} = \frac{1}{(1+z)H(z)}\,. 
\end{equation}
We also introduced
\be
\mathcal{A} = \int\, p(x) F(x)\; \mathrm{d}^2x\,.
\ee

In recent numerical simulations, e.g. the 'Illustris' simulation~\cite{2015MNRAS.454.2770T}, authors often determine the density of objects with fixed velocity dispersion and not with fixed mass. This is also what we shall use below. Furthermore, galaxies are not point masses but extended objects. If the matter distribution would be spherically symmetric, the exterior metric however would be Schwarzschild (Birkoff's theorem) and our Schwarzschild formula could be used. This is of course not truly the case and so our treatment gives only an order of magnitude estimation for the optical depth and the rotation probability.

We now use the fact that the mass enclosed inside the Einstein radius can be related to the velocity dispersion of the galaxy by (see e.g. Section II of \cite{Oguri:2019fix})
\begin{equation}
\label{eq:Mv}
M= \frac{4\pi^2}{G} \sigma_v^4 \frac{ \DL \DLS}{\DS}\,, 
\end{equation}
and we recall that comoving and physical number density are simply related by a factor of redshift cube 
\begin{equation}
n^{\text{phys}}=(1+z)^3 n^{\text{com}}\,.
\end{equation}
Hence (neglecting hereafter the label {'com'}) 
\begin{align}
n^{\text{phys}}(z,M) \mathrm{d}M &= (1+z)^3n(z, \sigma_v) \mathrm{d}\sigma_v\,.
\label{eq:nsigma}
\end{align}
The optical depth (\ref{eq:tau}) can be rewritten as 
\be\label{TauFin}
\tau(\Theta, \zS)=\frac{\mathcal{A}}{\Theta}\mathcal{J}(\zS)\,,
\ee
with
\be\label{JJ}
\mathcal{J}(\zS)=\int_{0}^{\zS} \frac{(1+z)^2}{H(z)}\left(\frac{\DL \DLS}{\DS}\right)^2 n(\sigma_v, z) \sigma_v^4 \mathrm{d}\sigma_v \mathrm{d}z\,.
\ee
Once the optical depth is known, the probability for a ray to reach the observer, while undergoing a scaled rotation larger than $\Theta$ can be computed as 
\begin{equation}
P(>\Theta, \zS) = 1 - \exp(- \tau( \Theta , \zS))\,,
\end{equation}
and the PDF for the variable $\Theta$ is
\begin{equation}
p(\Theta, \zS) = -\frac{\dd P(>\Theta, \zS)}{\dd\Theta} =- \frac{\partial \tau}{\partial \Theta} e^{- \tau} 
=
\frac{\mathcal A}{\Theta^2} \mathcal{J}(\zS) 
\exp\left( -\frac{\mathcal A}{\Theta} \mathcal{J}(\zS)\right)\,.
\end{equation}
While  $p(\Theta, \zS)$ is regular for $\Theta \ra 0$, it scales as 
$\Theta^{-2}$  for  $\Theta \ra \infty$, i.e. for $b^2\ra 0$. For this reason 
$\langle \Theta \rangle$ and actually $\langle \Theta^n \rangle$ for all $n\geq 1$ diverge at large $\Theta$.  But we know that our approximation is only valid for 
$$
b^2 > R^2_{\textrm{E}}=  \frac{4m \DLS \DL}{\DS}\, .
$$
We therefore need to introduce an upper cutoff in $\Theta$ given by
\begin{equation}
\Theta < \Theta_{\max} = \Theta(b=R_\textrm{E})={F}(\alphaS,z_s)=(2-\varepS^2)\vert \sin(2\alphaS)\vert  \,.
\end{equation}
This requirement of a cutoff is due to our simple lens model and could be removed with a more realistic modelling of the lens. However, it is not particularly relevant for the discussion in the present paper.

\subsection{Analytical results in a simple toy model}

To get some physical insight, we first consider a toy model for the comoving number density of galaxies: we use fits to observations at $z=0$ from \cite{Bernardi_2010}, see \ref{AppTau}, and we neglect the evolution of galaxies with redshift,  $n(z, \sigma_v)=n(\sigma_v)$. We use the following parametrization to fit $n(\sigma_v)$
\begin{align}
\label{Bernardi}
n(\sigma_v) &= 
\phi^\star 
\left( \frac{\sigma_v}{\Sigma_\star}\right)^{\alpha_B} 
\exp\left(- \left(\frac{\sigma_v}{\Sigma_\star}\right)^{\beta_B}\right) \frac{1}{\Gamma\left(\frac{\alpha_B}{\beta_B}\right)} \frac{\beta_B}{\sigma_v}\,,
\end{align}
{where the values of the numerical coefficients highly depend on the sample under study and are listed in \ref{AppTau} for the SDSS catalogue}. This is a generalization of the Schechter function (see e.g.~\cite{Peacock:1999}) commonly used to fit to the luminosity function of galaxies.  Below, to derive more realistic predictions, we will include the effect of evolution using results from the numerical simulations of \cite{2015MNRAS.454.2770T} (the 'Illustris simulation'), see \ref{Tau2} for details. 

The advantage of working with (\ref{Bernardi}) is that the integral in eq.~\eqref{eq:tau} can be factorised into two integrals, over $z$ and $\sigma_v$ and computed analytically. 
The integral over $\sigma_v$ is
\begin{equation}
\int_0^\infty\, \sigma_v^4 n(\sigma_v)\; \mathrm{d}\sigma_v
=
(\Sigma_\star)^4 \phi^\star \frac{\Gamma\left( \frac{4+\alpha_B}{\beta_B} \right)}{\Gamma\left( \frac{\alpha_B}{\beta_B} \right)}\,.
\end{equation}
The $z$-dependent integral is
\begin{equation}
\label{eq:Izs}
\mathcal{I}(z_s)=
\int_0^{z_S} \;\mathrm{d}z\,\frac{(1+z)^2}{H(z)} 
\left(\frac{\DL \DLS}{\DS}\right)^2
=\int_0^{z_S}\;\mathrm{d}z
 \,\frac{1}{H(z)}
 \left(\frac{\chi(0,z) \chi(z,\zS)}{\chi(0,\zS)}\right)^2\,,
\end{equation}
where we introduced the comoving distance from $z_1$ to $z_2$ as $\chi(z_1,z_2)$.
Combining everything, we get for the function $\mathcal{J}$ in (\ref{JJ}) 
\begin{equation}
\mathcal{J}(\zS) \equiv 
\frac{16\pi^3 \sigma_\star^4 \phi^\star }{H_0^3} 
\frac{\Gamma\left( \frac{4+\alpha_B}{\beta_B} \right)}
{\Gamma\left( \frac{\alpha_B}{\beta_B} \right)} 
H_0^3 \mathcal{I}(\zS)\,.
\end{equation}
Note that the function intergrated in Eq.~\eqref{eq:Izs} scales as $H_0^{-3}$, as $\chi(z_1,z_2)\sim1/H_0$ and $\mathcal I \sim \chi^2/H  \sim 1/H_0^3$, while $\phi^\star$ has the units of $H_0^3$. Hence, we introduced some $H_0^3$ factors to stress that all the quantities appearing are dimensionless.

In Section~\ref{sec:numres} (cf. Fig~\ref{fig:toyVSill}) we will show a comparison between the toy model, the  evolving Illustris model and the Illustris model neglecting evolution in the expression for the density of galaxies $n$.

\subsection{Numerical results} \label{sec:numres}
To derive more realistic predictions, we refine the way we model the galaxy comoving number density, including the effect of redshift evolution. We use results from the numerical simulations of \cite{2015MNRAS.454.2770T}, the so called 'Illustris simulations'. They have parametrised the galaxy number density as a function of the velocity dispersion and of redshift, $n(\si_v,z)$, see \ref{Tau2} for details. Using this 'Illustris model' we have to determine the optical depth (\ref{TauFin}) and the corresponding probability numerically. 
\begin{figure}[!ht]
        \centering
        \includegraphics[width=0.49\columnwidth]{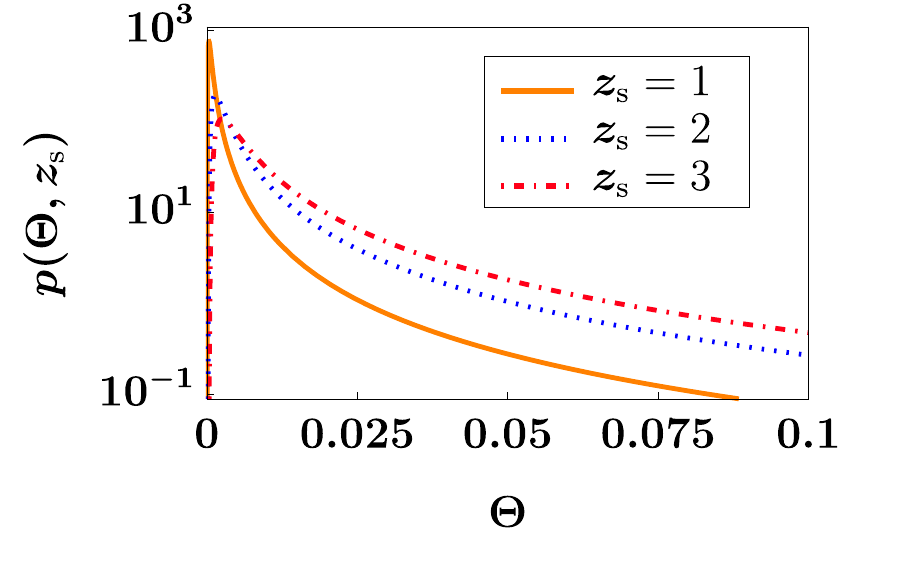}
        \includegraphics[width=0.49\columnwidth]{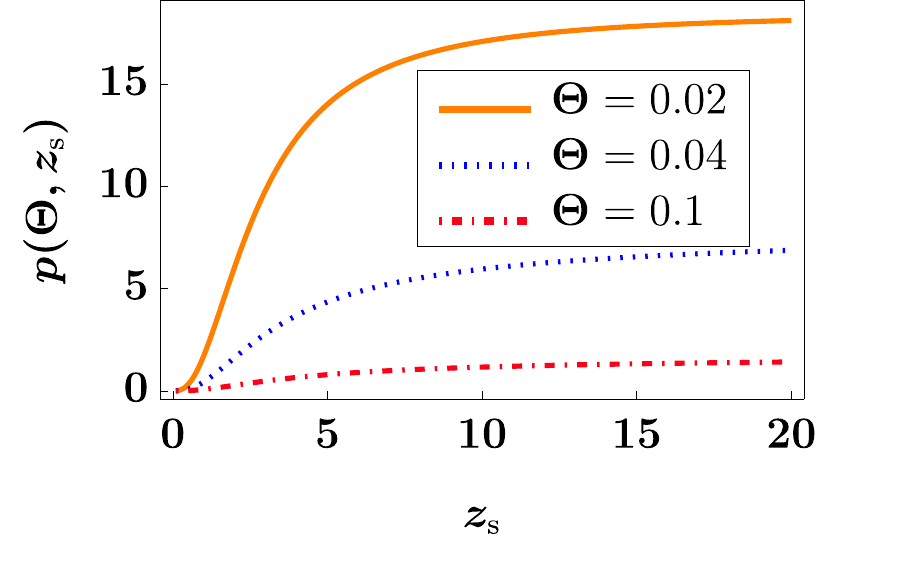}
    \caption{\label{fig:PDFOZS} PDF for the scaled rotation $\Theta$ as function of $\Theta$ for different source redshifts (left panel) and as function of redshift for different values of $\Theta$ (right panel). We used the Illustris model for the galaxy distribution $n(\si_v,z)$.}
\end{figure}

The only quantity left to compute is the normalization constant $\mathcal A$, coming from the average over the parameters of the source $x = (\alphaS, \varepS)$. 
To compute the value of $\mathcal A$, we assume that the distribution of the angle $\alpha_s$ is flat, and for the distribution of intrinsic eccentricity we use results of \cite{chen2016}. The resulting distribution of intrinsic parameters is given by 
\begin{equation}
p(\alphaS, \varepS) = C \exp 
\left(- \frac{(\varepS-\varepsilon_0)^2}{2\sigep^2}\right)\,,
\end{equation}
with $\varepsilon_0=3/10$ and $\sigep=5/100$. 
The constant $C$ is determined by the normalization condition,
\begin{equation}
\int_0^1\; \mathrm{d}\varepS
\int_0^{\frac \pi 2}\; \mathrm{d}\alphaS\, p(\alphaS, \varepS) =1\,. 
\end{equation}
In Fig.\,\ref{fig:PDFOZS} we show the resulting PDF of the scaled rotation $\Theta$ as function of redshift for different fixed $\Theta$ and as function of $\Theta$ at different $z_s$. As expected, the probability density increases with redshift and it decreases as the rotation $\Theta$ increases.  

\begin{figure}[!ht]
        \centering
        \includegraphics[width=0.8\columnwidth]{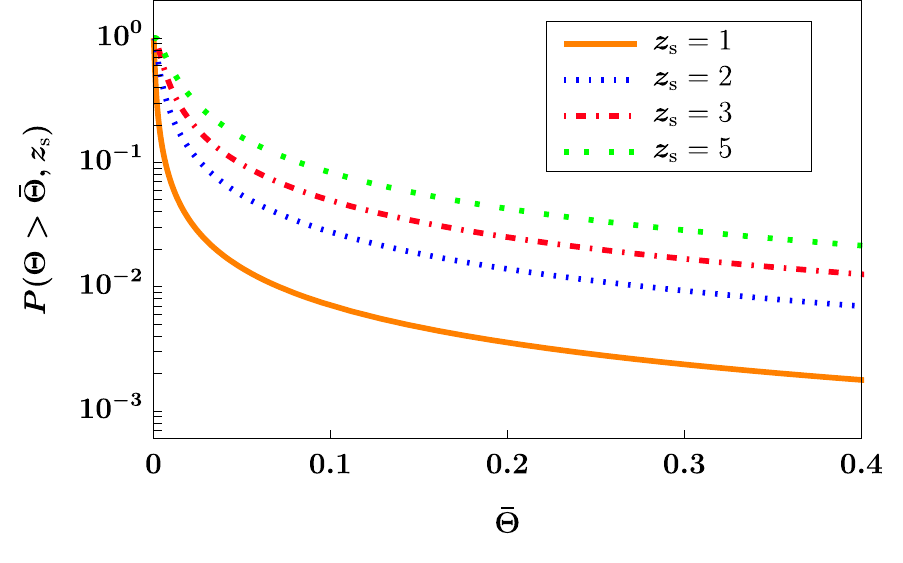}
    \caption{\label{fig:PP} Probability of being lensed with rescaled rotation larger than $\bar{\Theta}$, as function of $\bar{ \Theta}$ for different $\zS$. We used the Illustris model for the galaxy distribution $n(\si_v,z)$.}
\end{figure}

 As already mentioned, at large values of $\Theta$, for all redshift the scaling is $p(\zS, \Theta)\propto \Theta^{-2}$. It follows that mean and variance of the distribution will be both dependent on the value of a cut off  used to regularise the integral over $\Theta$. Hence a better observable to look at to have an idea of the size of the effect, is the probability to get a rotation bigger than a fixed $\bar\Theta$ for a source at a given redshift. Indeed the PDF is a well-defined function, whose integral converges, hence this result will not require a cut-off. Results are shown in Fig.\,\ref{fig:PP}. For a source at redshift $\zS=2$ the probability of being lensed with rescaled rotation larger than $0.05$ radian (about $3^o$) is 10\%. 

\begin{figure}[!ht]
        \centering
        \includegraphics[width=0.8\columnwidth]{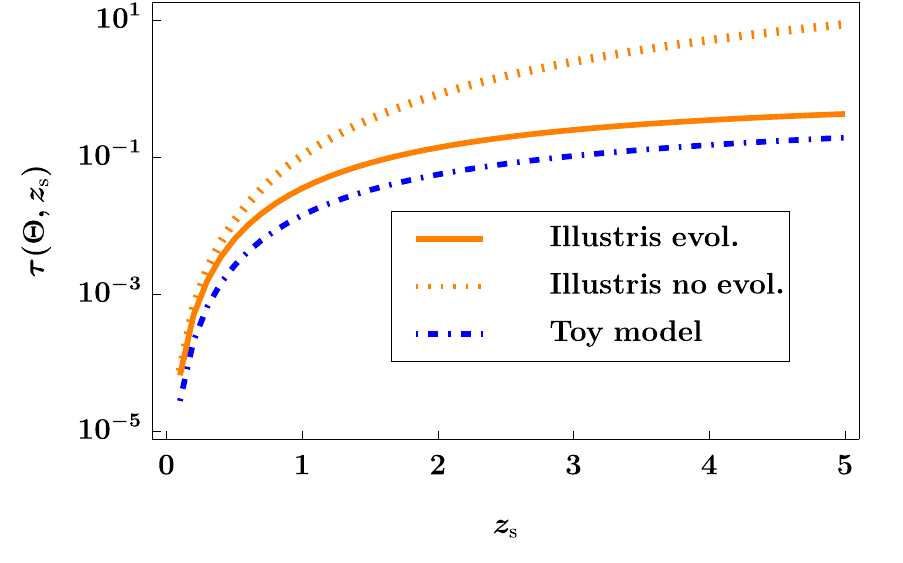}
    \caption{\label{fig:toyVSill} Optical depth, using the toy model and the Illustris simulation. Both curves are obtained for $\mathcal O = 0.02$. Comparing the two Illustris curves, it is clear that the evolution of the galaxy number density is a very important ingredient for our results. }
\end{figure}

In Fig.~\ref{fig:toyVSill}, we show the difference for the optical depth using the toy model of the previous Section, the Illustris simulation and the Illustris simulation at $z=0$ but neglecting the evolution of the density of galaxies (hence only considering the dilution due to the expansion).

\section{Conclusions}\label{sec:con}

In this article, we discuss the use of the polarisation direction as a proxy for the intrinsic structural direction of a radio galaxy. For low frequency radio galaxies, the dominant source of linear polarisation is  synchrotron radiation from electrons moving in the magnetic field of the galaxy. In this case, the polarisation vector is expected to be aligned with the galaxy semi-minor axis at emission.  It follows that a non-vanishing  angle between these two vectors, if measured,  is a signature of lensing. Making use of the standard Jacobi formalism for weak lensing, we derive analytical expressions for this angle, as function of the gravitational potential along the line of sight and of intrinsic parameters characterizing the galaxy. 

We specialise our results to the case of a Schwarzschild lens. We show that in realistic scenarios this angle is quite sizeable, e.g. for a lens with $M\sim10^{11}M_\odot$ located at a distance of $\DL=10\; \mathrm{Mpc}$ with an impact parameter $b=10\;\mathrm{kpc}$, the rotation angle is $\delta \al\sim 0.38^\circ$ for $\vep\sim 1/2$. For significantly smaller eccentricity or larger galaxies,  $\delta \al$ can also be much larger. We then consider a statistical distribution of lenses (foreground galaxies). For a realistic galaxy distribution, we find that the probability of being lensed with a scaled angle $\Theta=\vert \delta\alpha \vert \ \varepO^{2}>0.4\, \textrm{rad}$, which corresponds to about $23^o$ is larger than 1\% for sources at a redshift larger than $\zS=2$. {Or the probability for lensing rotation by $6^\circ/\vep^2$ by a lens at $z\leq 2$ is $\sim$ 3\%, while at $z=1$ it is smaller than 1\%.}

The goal of this paper is to propose a new method to measure the integrated gravitational potential, exploiting polarisation information, that will come for free in forthcoming radio surveys. We show that the observed polarisation provides information on both the unlensed source orientation and the gravitational potential along the line of sight, and we derive predictions in  simple  scenarios to broadly assess the potential of our method. More realistic predictions can be derived considering a cosmological context and a cosmological distribution of the gravitational potential. 

The precision with which the polarisation can be determined (which takes into account both intrinsic scattering and measurement error) is between 1 and 10 degrees depending on the class of galaxies under study see \cite{Stil:2008ew} and \cite{Thomas:2017ewl}. This means that a suffiently large sample is needed to reach the required precision and be able to extract the rotation effect discussed in this paper. Moreover, in order to apply our analysis to real data, we will need also to take into account the rotation of the polarisation plane due to Faraday rotation. Faraday rotation by the stochastic magnetic field in and between galaxies results in a frequency-dependent depolarisation: as linearly polarised emission travels through a birefringent magnetised media (e.g. the intergalactic medium), a difference in the phase velocity occurs for the right and left circularly polarizated radiation. This manifests itself as a wavelength-dependent rotation of the polarisation angle, see e.g. \cite{Kronberg:1993vk,Mahatma_2020}. However, correcting for Faraday rotation should be possible by using multi-frequency observations. We will investigate in detail these aspects, and other issues related to observations, in a future work.

Even though our approach depends in principle on the intrinsic eccentricity and orientation of the lens, at first order, we can replace the intrinsic eccentricity by the observed one. However, once we consider a stochastic gravitational potential, there will no longer exist a fixed principal direction of the shear. Also this direction will be a random variable and due to statistical isotropy we expect $\langle\delta\alpha\rangle=0$. However, its variance will be non-vanishing as well as its angular power spectrum. We expect the redshift dependent $\de\al$ power spectrum to be directly related to the shear power spectrum.~\cite{prep}

We observe that our present analysis is only valid within the framework weak lensing, small shear. It might be interesting to study it also for strong lenses in the presence of several images and a realistic lens model.

\section*{Acknowledgement}
We thank Stefano Camera, Jean-Pierre Eckmann, Pierre Fleury, Martin Kunz and Francesca Lepori for helpful discussions. This work is supported by the Swiss National Science Foundation.

\vspace{3cm}

\appendix
\section{Images are Lie transported}
In this appendix we show that in the weak lensing approximation images are actually Lie transported. We consider two neighboring lightlike geodesics with velocity vectors $\boldsymbol{k}$ and $\boldsymbol{k^\prime}$
connected by a 'geodesic deviation vector'  $\boldsymbol\xi$.  The geodesic deviation equation
then reads
\begin{equation}\label{e:geodev}
\ddot \xi^\al =  -R^\al_{\beta\mu\nu}k^\beta \xi^\mu k^\nu \,.
\end{equation}
We now show that this implies that $\boldsymbol\xi$ is Lie transported along $\boldsymbol k$, i.e. $L_{\boldsymbol k}\boldsymbol\xi=0$.
This is easiest seen in coordinate free notation where 
\begin{equation}
\ddot{\boldsymbol\xi}=\boldsymbol{\nabla_k\nabla_k\xi}\,,
\end{equation}
and 
\begin{equation}
-R^\al_{\beta\mu\nu}k^\beta \xi^\mu k^\nu = -(\boldsymbol R(\boldsymbol\xi,\boldsymbol k)\boldsymbol k)^\al = (\boldsymbol{\nabla_k\nabla_\xi-\nabla_\xi\nabla_k})\boldsymbol k=\boldsymbol{\nabla_k\nabla_\xi k} \,.
\end{equation}
Here we have used $\boldsymbol{\nabla_kk}=0$. Hence (\ref{e:geodev}) implies 
\begin{equation} 
0=\boldsymbol{\nabla_k}(\boldsymbol{\nabla_k\xi}-\boldsymbol{\nabla_\xi k})= \boldsymbol{\nabla_k}([\boldsymbol k,\boldsymbol\xi])=\boldsymbol{\nabla_k}\left(L_{\boldsymbol k}\boldsymbol\xi\right)\,,
\end{equation}
where $L_{\boldsymbol k}$ denotes the Lie derivative in direction $\boldsymbol k$.  Hence $L_{\boldsymbol k}\boldsymbol\xi$ is constant along the geodesics.

However, in the source plane, $\la=\la_{\rm{in}}$, we can always choose $\boldsymbol\xi$ to denote a coordinate direction, $\boldsymbol\xi=\boldsymbol\pp_s$ and $\boldsymbol k=\boldsymbol\pp_\la$ so that at $\la_{\rm{in}}$ we have $$[\boldsymbol k,\boldsymbol\xi]=[\boldsymbol\pp_\la,\boldsymbol\pp_s]=0\,.$$ The constancy of $L_{\boldsymbol k}\boldsymbol\xi$ therefore implies $L_{\boldsymbol k}\boldsymbol\xi=0$ along the photon geodesic.

\section{Decomposition of a two-dimensional square matrix \label{sec:appmatrix}}
In this Section, we explain how one can obtain the decomposition given by Eq.\eqref{eq:jacobicompact} for a non-singular $2\times 2$ matrix with positive eigenvalues. Let 
\begin{equation}
    A = 
    \begin{pmatrix}
    a_{11} & a_{12} \\
    a_{21} & a_{22}
    \end{pmatrix}
\end{equation}
be an arbitrary $2\times 2$ matrix, and
\begin{equation}
R(\chi) =
\begin{pmatrix}
\cos \chi &  \sin \chi \\ 
-\sin \chi& \cos \chi 
\end{pmatrix}
\end{equation}
be a rotation.
If $a_{12}\neq a_{21}$, it is easy to check that the matrix $\tilde A = R(\psi) A$ is symmetric if $\psi$ is chosen such that
\begin{equation}
(a_{12}-a_{21}) \cos \psi + (a_{12} + a_{22}) \sin(\psi) =0\,,
\end{equation}
which can be solved for $\psi$ (the case $a_{12}=a_{21}$ is trivial with $\psi=0$ as the matrix is already symmetric). Using the spectral decomposition theorem, the new matrix $\tilde A$ can be diagonalised with a rotation matrix, $$R(-\phi)\tilde A R(\phi)=
\begin{pmatrix}
a_+ &  0 \\ 
0& a_- 
\end{pmatrix} \,.$$ 
As the eigenvalues $a_\pm >0$, one can define $D=\sqrt{a_+a_-}\; >0$ and $\gamma =\log(a_+/a_-)/2$ such that
\begin{equation}
    D e^{\pm\gamma} = a_\pm\,.
\end{equation}
Combining these results leads to the decomposition Eq.\eqref{eq:jacobicompact}.

\section{Rotation of an ellipse due to lensing} \label{sec:appmaps}
When the image of the sources undergoes weak-lensing, it is deformed. We describe here the method to compute quantitatively this deformation.  At the source position, on the Sachs screen,  the projected ellipse has eccentricity $\varepS$, and we normalise the axis such that the minor-axis has length $1$ and the semi-major axis has length $(1-\varepS^2)^{-1/2}$. The semi-major axis forms an angle $\alphaS$ with the first Sachs vector. Denoting by $\chiS$ a vector on the Screen space at the source position, such an ellipse is defined by the condition
\begin{equation}
\label{eq:condellipseS}
\chiS^{\mathrm{T}} \AS \chiS = 1\,,
\end{equation}
with 
\begin{equation}
\AS = R(-\alphaS) 
\begin{pmatrix}
1-\varepS^2 &0 \\ 
0 &1 
\end{pmatrix}
 R(\alphaS)\,.
\end{equation}
Indeed, one can check explicitly that the vectors $\boldsymbol{v}_1 = (1-\varepS^2)^{-1/2} (\cos \alphaS, \sin \alphaS)$ and $\boldsymbol{v}_2 =  (\cos (\alphaS+\pi/2) , \sin (\alphaS + \pi/2))$ corresponding to the principal axis satisfy Eq.~\eqref{eq:condellipseS}.

The next step is to determine the condition defining the image shape at the observer. We use Eq.~\eqref{eq:thetatoksi} relating the observed angle to the position on the Sachs screen at the source, and replace in \eqref{eq:thetatoksi} the vector at the source position $\chiS =\jacD\thetaO$ in Eq.~\eqref{eq:condellipseS}. This gives 
\begin{equation}\label{eq:condellipseO}
{\thetaO}^{\mathrm{T}} A_\textrm{o}
{\thetaO}=1\,,
\end{equation}
with
\bea
A_\textrm{o} &=& \jacD^{\mathrm{T}} A_\textrm s \jacD\\
&=& D_s^2R(-\phi)\left(\begin{array}{cc}e^{-\ga} & 0\\ 0 &e^\ga\end{array}\right)  R(\psi+\phi-\al_s)\begin{pmatrix}
1-\varepS^2 &0 \\  0 &1 \end{pmatrix}
\times \nonumber \\
&& \qquad
 R(-\psi-\phi+\alpha_s)R(-\phi)\left(\begin{array}{cc}e^{-\ga} & 0\\ 0 &e^\ga\end{array}\right)R(\phi) \,.
\eea
For the second equal sign we inserted \eqref{eq:jacobicompact}.
With $A_s$ also $A_o$ is symmetric and it can be diagonalized. Hence
\be
A_\textrm{o} = N R(-\alphaO) 
\begin{pmatrix}
1-\varepO^2 &0 \\ 
0 &1 
\end{pmatrix}
 R(\alphaO)\,. 
\ee
The normalization factor $N$ is defined as the larger of the two eigenvalues of $A_\textrm{o}$ and $\varepO$ is fixed by requiring that $(1-\varepO^2)$ is the ratio of the two eigenvalues.
We determined the eigenvalues and eigenvectors of the $2\times 2$ matrix $A_\textrm{o}$ with \emph{Mathematica}. The full result is not very illuminating, but we then use that $\psi$ and $\ga$ are small and expand our result to first order in the small shear $\ga$ and lensing rotation $\psi$. The angle  of the eigenvector with the smaller eigenvalue to the $x$-axis  determines $\alphaO$.

 The final results are
\begin{equation}
\label{eq:StoO}
\varepO = 
\varepS  + \gamma \cos \left[2 (\alphaS-\phi)\right]\frac{2 (1-\varepS^2) 
}{\varepS}\,, \quad
\alphaO = \alphaS - \psi - \gamma \sin \left[2 (\alphaS-\phi)\right] \frac{2-\varepS^2}{\varepS^2}\,.
\end{equation}
These expressions can be understood qualitatively. First, the quantity $\psi$ quantifies a global rotation counterclockwise \emph{from} the observer screen \emph{to} the source screen. However, we are now considering how the image shape gets deformed from the source to the observer, and this explains the negative sign in front of $\psi$ in the expression for $\alphaO$ in (\ref{eq:StoO}).
Then, if one rotates (counterclockwise) the two Sachs vectors by an angle $\chi$ to define a new Sachs basis $(\tilde \bolde_1, \tilde \bolde_2)$, the angles $\alphaS$ and $\phi$ are modified as $\tilde \alphaS = \alphaS - \chi$ and $\tilde \phi = \phi - \chi$. However, the rotation of the semi-major axis and the eccentricity of the ellipse at the observer should not depend on the angle $\chi$. This explains why only the difference $\alphaS-\phi$ appears. Finally, the quantities $\alphaS$ and $\phi$ are spin-2 variables in the sense that the physics is invariant if $\phi$ is replaced by $\phi+\pi$, which is reflected in the factor $2$ in the trigonometric functions in Eq.~\eqref{eq:StoO}. 

There are two interesting cases. First, let us assume $\alphaS=\phi$ and consider an ellipse at the source position with the semi-major axis aligned with the first Sachs vector. Using the relation between the observed and the source vector given by Eq.~\eqref{eq:thetatoksi}, we see that the semi-major axis at the observed position is elongated and the minor-axis is contracted. This implies that the eccentricity is amplified at the observed position, hence $\varepO > \varepS$ which is consistent with the result shown above using $\varepS<1$.

Secondly, if $\alphaS-\phi = \pi/4$, the eccentricity of the ellipse is not modified and the ellipse undergoes a net rotation.

Finally, we note that these formulas do not work if $\varepS\approx 0$. Indeed, in this case the ellipse is almost a circle, and a small shear can have a big impact on the shape of the ellipse, hence the perturbative expansion breaks down. 

\section{Orthonormal frame and spatial vectors for the Schwarzschild lens\label{sec:apportho}}

We detail here the spatial description of the setup of Section ~\ref{sec:ss}. The 4-velocity of an observer at a distance $r$ from the lens is given by 
\begin{equation}
\boldsymbol{u} =\frac{1}{\sqrt{ 1- \frac{2m}{r}}}\pp_t\,,
\end{equation}
and the induced metric on spatial hypersurfaces is
\begin{equation}
\boldsymbol h = \boldsymbol g + \boldsymbol u \otimes \boldsymbol u
=
 \left( 1- \frac{2m}{r} \right)^{-1}\dd r^2 + r^2\; \dd \theta^2 + r^2 \sin^2 \theta\; \dd \varphi^2\,,
\end{equation}
or in components
\begin{equation}
h_{\mu \nu} = g_{\mu \nu} + u_\mu u_\nu, \quad
h^{\mu \nu} = g^{\mu \nu} + u^\mu u^\nu,
\quad 
h^\mu_\nu = \delta^\mu_\nu +u^\mu u_\nu \,.
\end{equation}
Here 
$h^\mu_\nu$ is the projector onto the spacelike hypersurface. Applying it on the momentum $\boldsymbol k$ of the photon, we extract its spatial part $\boldsymbol{k}^{\rm{sp.}}$,
\begin{equation}
\boldsymbol{k}^{\rm{sp.}} = \dot r\pp_r +\frac{L}{r^2}\pp_\varphi\,,
\end{equation}
with
\begin{equation}
\dot r = \pm \sqrt{1 - \frac{L^2}{r^2} + \frac{2mL^2}{r^3}} \,,
\end{equation}
where $L$ is the angular momentum of the photon. We can now compute the spatial angle $\zeta$ between the vector $-\boldsymbol{k}_{\rm{sp.}}$ and $\boldsymbol{a} \equiv  \partial_r$, which is the angle between the incoming photon and the position of the lens, measured by an observed. One can use the usual formula for the scalar product to get
\begin{equation}
\cos \zeta =\frac{
a^\mu h_{\mu \nu} k^\nu
}
{
\sqrt{a^\mu h_{\mu \nu} a^\nu} 
\sqrt{k^\mu h_{\mu \nu} k^\nu}
}\,,
\end{equation}
where we can use $\boldsymbol k$ as $h_{\mu \nu}k^\mu = h_{\mu \nu} k_{\rm{sp.}}^\nu$. This leads to the expression for $\sin \zeta$ given by Eq.~\eqref{eq:sinzeta}.

It is also possible to introduce an orthonormal frame defining the set of four forms
\begin{equation}
\boldsymbol \sigma^a = \tensor{e}{^a_\mu} \mathrm{d}x^\mu\,,
\end{equation}
and the corresponding set of four vectors
\begin{equation}
\boldsymbol \sigma_a = \tensor{e}{^\mu_a} \partial_\mu\,,
\end{equation}
where $ \tensor{e}{^\mu_b} \tensor{e}{^b_\nu}= \delta^\mu_\nu$ and $g_{\mu \nu}\tensor{e}{^\mu_a}\tensor{e}{^\nu_b}=\eta_{ab}$. In this orthonormal basis, the metric simply reads $\boldsymbol g = \eta_{ab}\boldsymbol\sigma^a \otimes \boldsymbol\sigma^b$. The matrix $\tensor{e}{^a_\mu}$ is easily found and is diagonal with components
\begin{equation}
\tensor{e}{^0_t} = \sqrt{1- \frac{2m}{r}}\,,\quad
\tensor{e}{^1_r} = \left(\sqrt{1- \frac{2m}{r}}\right)^{-1}\,,\quad
\tensor{e}{^2_\theta} = r\,,\quad
\tensor{e}{^3_\varphi} =r \sin \theta \,.
\end{equation}
Given a 4-vector $\boldsymbol V = V^\mu \partial_\mu$, its component in this orthonormal frame are given by
\begin{equation}
\boldsymbol V = V^a \boldsymbol\sigma_a, \quad V^a = V^\mu \tensor{e}{^a_\mu}.
\end{equation}

We can apply this formalism to the Sachs basis $\bolde_1$ and $\bolde_2$ to compute their components in the orthonormal basis. As expected, their $0$-th component vanish. Seen as spatial vectors, one can check by inspection that $(\boldsymbol\sigma_1, \boldsymbol\sigma_2, \boldsymbol\sigma_3)$ forms a direct frame. This also holds for $(\bolde_1, \bolde_2, \hat{\boldsymbol{k}}^{\rm{sp.}} )$, where the last vector is the normalised spatial part of the momentum of the photon. This is in agreement with the intuition motivated by Fig.~\ref{f:ss-sachs} and this also shows that, for an observer knowing the location of the lens and detecting an incoming photon with spatial 4-momentum $\boldsymbol{k}^{\rm{sp.}}$, the Sachs basis can be unambiguously defined: the first basis vector $\bolde_1$ is normal the orbital plane, and the second basis vector $\bolde_2$ is defined by the condition of the triplet to be orthogonal and direct. There is one more ambiguity as one can define $\tilde \bolde_A = - \bolde_A$ for $A=1,2$. However, as the ellipse considered on the Sachs screen is symmetrical under a rotation by $\pi$, this ambiguity does not change the physical observables.

\section{Measuring the orientation of an ellipse in the sky}\label{ap:error}

Typically we can measure the observed angular radius $r(\vph)$ of an image in the sky with some error. Assuming this image to be an ellipse with semi-major axis $a$ at an angle $\alpha$ to our (arbitrary) $x$-axis and with eccentricity $\vep$, we model it as the function  
\be
r(\vph) = a\sqrt{1-\vep^2\sin^2(\vph-\al)}\,.
\ee

Let us consider in more detail how we determine the orientation $\al$ of the ellipse. We measure distances $r_i$ from some center in direction $\vph_i$ from some arbitrary $x$-axis. We neglect the error in the direction but assume some error $\si_r$ for the radial distance from the center. For simplicity we assume the same accuracy for all measurements $r_i$. We want to fit the points $(r_i,\vph_i)$ with an ellipse given by
\be
r(\vph_i) = a\sqrt{1-\vep^2\sin^2(\vph_i-\al)}\,,
\ee
by minimizing $\chi^2(\al)$ defined as
\be
\chi^2(\al)=\frac{1}{N}\sum_{i=1}^N\frac{(r_i-r(\vph_i))^2}{\si_r^2}\,.
\ee
In principle one would of course estimate all $(a,\al,\vep)$ in this way, but here we are only interested in the behavior of the error in $\al$ while keeping $a$ and $\vep$ fixed.
Hence best estimate, $\hat\al$, for the principal direction $\al$ is the solution of
\be\label{e:chi-min}
0 =\frac{\pp\chi^2(\al)}{\pp\al} =\frac{1}{N}\sum_{i=1}^N\frac{1-r_i/r(\vph_i)}{\si_r^2}a^2\vep^2\sin(2(\vph_i-\al)) \,.
\ee

   \begin{figure}[!ht]
        \begin{center}
        \includegraphics[width=0.8\columnwidth]{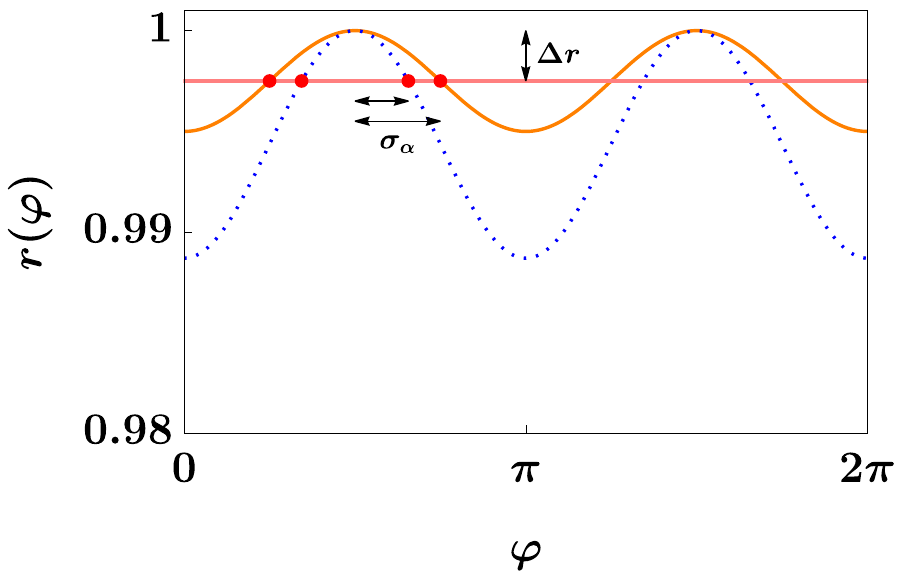}
        \caption{\label{fig:ep} We show how an uncertainly in the size of the ellipse (assumed to be constant, indicated by the red horizontal line) translates into an uncertainty of the orientation of the ellipse given by $\De\al$ indicated by the two red dots on the lines $r(\vph)$. The full orange line is for $\vep=0.1$ while the dashed blue line has $\vep=0.15$.}
        \end{center}
    \end{figure}
The error is given by the second derivative,
\bea
\si_\al^{-2} =\left.\frac{\pp^2\chi^2}{\pp\al^2}\right|_{\al=\hat\al} &=& \frac{1}{N\si_r^2}\sum_{i=1}^N\Bigg[\left(\frac{r_i}{r(\vph_i)}-1\right)a^2\vep^2\cos(2(\vph_i-\al)) \\  &&  \qquad\qquad
+\frac{r_i}{r(\vph)^3}a^4\vep^4\sin^2(2((\vph_i-\al))\Bigg] \,.
\eea
At $\al=\hat\al$ we have $r_i\simeq r(\vph_i)$. Inserting this above we obtain
\be
\si_\al^{-2} =\frac{a^2\vep^4}{\si^2_r}\frac{1}{N}\sum_i\frac{\sin^2(2((\vph_i-\al))}{1-\vep^2\sin^2(\vph_i-\al)} \,.
\ee
Replacing the sum over the angles $\vph_i$ by an integral, we obtain
\bea
\frac{1}{N}\sum_i\frac{\sin^2(2((\vph_i-\al))}{1-\vep^2\sin^2(\vph_i-\al)} &\simeq&
\frac{1}{2\pi}\int_0^{2\pi}\,\frac{\sin^2(2\vph)}{1-\vep^2\sin^2\vph}\;\dd\vph
 ~~~~  \nonumber \\  &&
 \hspace*{-3cm} =\frac{2  \left(2-\vep^2-2 \sqrt{1-\vep^2}\right)}{\vep^4}\simeq \frac{1}{2} + \frac{\vep^2}{4} +\frac{5\vep^4}{32} + {\cal O}(\vep^6) \,.
\eea
To lowest order in $\vep$ we therefore have
\be
\si_\al \simeq\sqrt{2}\frac{\si_r}{a}\vep^{-2} \,.
\ee
For this reason we consider the variable ${\cal O}=\vert \de\al \vert \vep^2$ as our observable. This variable is well behaved in the limit $\vep\rightarrow 0$, while the error of $\de\al$ diverges in this limit. 
In Fig.~\ref{fig:ep} the function $r(\vph)$ and the inferred error in $\al$, $\si_\al$, are shown for two values of $\vep$.

\section{Statistical approach: derivations and fitting formulas}
\subsection{Optical depth}

Here we summarise the key steps to compute the optical depth and the PDF of a given observable. Assume we have a trajectory through a \emph{tube} from the source to the observer filled with targets. For a given trajectory, we can associate an observable $\Theta$, which is influenced by the targets through a relationship $\Theta(b)$, where $b$ is the impact parameter of the trajectory around a given target. We assume that this relationship is monotonic and satisfies 
$\Theta(\infty)=0$. We want to compute the probability, for a given trajectory, for the observable to have a value bigger than, say $\Theta_0$ (and we call this a rotation). Let us consider a small portion of the tube, of physical length $\mathrm{d}r$ and of area $A$, and let $n$ be the density of targets. The probability of rotation by $\Theta_0$ or more is 
\begin{equation}
\mathrm{d}P = \frac{\sigma_0}{A} n A \mathrm{d}r = \sigma_0 n \mathrm{d}r\,,
\end{equation}
where $\sigma_0 = \sigma(\Theta_0)=\pi b_0^2$, where $b_0$ is defined by $\Theta(b_0)=\Theta_0$. If $\Theta(b)<\Theta_0$ for all values of $b$, we simple set $\si_0=0$. The probability of not being rotated by $\Theta_0$ or more in this portion of the tube is
\begin{equation}
\mathrm{d}P  = 1- \sigma_0 n \mathrm{d}r\,.
\end{equation}
If we denote by $P(r)$ the probability of reaching a distance $r$ without rotation by $\Theta_0$ or more, we can use the above relation to write
\begin{equation}
P(r+\mathrm{d}r)=P(r)( 1- \sigma_0 n \mathrm{d}r)\,.
\end{equation}
Solving this differential equation we find
\begin{equation}
P(\Theta>\Theta_0,r)=\exp\left(-\int_0^r \sigma(\Theta_0,y) n(y)\; \mathrm d y\right)
\equiv \exp (-\tau (\Theta_0,r))\,,
\end{equation}
where we have added the $\Theta$ dependence which is relevant in our context. The quantity $\tau (\Theta_0) $ is called the \emph{optical depth}. We want to compute the probability of rotation by $\Theta_0$ or more, which is given by $1-P$,
\begin{equation}
P(\Theta>\Theta_0,r) = 1-\exp ( -\tau (\Theta_0,r) )\,.
\end{equation}
Moreover, in our situation, we can parametrise the physical distance using the redshift, making use of
\begin{equation}
\mathrm{d}r = \frac{\mathrm d r}{\mathrm d z} \mathrm d z\,, \quad 
r \rightarrow \zS\,,
\end{equation}
and we can consider the presence of intrinsic parameters $x$ and replace
\begin{equation}
\sigma(\Theta_0,z) n(z) = \int\; \mathrm d x\, \sigma(\Theta_0,z,x) n(z,x)\,.
\end{equation}
Combining these results, we find
\begin{align}
P(\Theta>\Theta_0, \zS) &= 1- \exp(-\tau(\Theta_0, \zS)) \\
\tau(\Theta_0, \zS) &= \int\, p(\chi) n(z,x,M) \frac{\mathrm{d}r}{\mathrm{d}z} \sigma(\Theta_0, z, \zS, x, M)
\; \mathrm{d}z
\; \mathrm{d}M 
\; \mathrm{d}x\,,
\end{align}
where $M$ represents the mass of the lens, and $x$ the intrinsic parameters characterizing the image at emission, over which we integrate with the probability distribution $p(x)$. If we introduce the PDF of the variable $\Theta$, $p(\Theta,\zS)$ as
\begin{equation}
P(\Theta>\Theta_0, \zS) = \int_{\Theta_0}^\infty\, p(\Theta,\zS)\; \mathrm d \Theta\,,
\end{equation}
we obtain directly 
\begin{equation}
p(\Theta, \zS) = - \frac{\partial \tau}{\partial \Theta} e^{- \tau} \,.
\end{equation}

\subsection{The comoving number density of galaxies: a toy model}\label{AppTau}

In Section \ref{sec:stoch}, we first use a toy model for the comoving number density of galaxies (our lenses), useful to obtain analytical expressions and to gain physical insight: we assume the comoving number density to be only a function of the velocity dispersion (i.e. we neglect evolution of galaxies with redshift). For $n^{\text{com}}(\sigma_v)$  we use the fits to SDSS observations from \cite{Bernardi_2010}
\begin{align}
n^{\text{com}}(\sigma_v) &= 
\phi^\star 
\left( \frac{\sigma_v}{\Sigma_\star}\right)^{\alpha_B} 
\exp\left(- \left(\frac{\sigma_v}{\Sigma_\star}\right)^{\beta_B}\right) \frac{1}{\Gamma\left(\frac{\alpha_B}{\beta_B}\right)} \frac{\beta_B}{\sigma_v}\,. 
\end{align}
This is a generalization of the Schechter function commonly used to fit to the luminosity function. The fit coefficients strongly depend on the sample chosen. The mean values found in \cite{Bernardi_2010} fitting the entire sample are
\begin{align*}
\Sigma^\star &= 113.78\mathrm{km\times s^{-1}}\approx 0.0038\,, \\
\phi^\star &= 1.65\times 10^9 H_0^3\,, \\
\alpha_B &= 0.94\, \quad \textrm{and}  \\
\beta_B &= 1.85 \,.
\end{align*}

\subsection{The comoving number density of galaxies  including redshift evolution}\label{Tau2}
To obtain more realistic predictions, one has to consider the fact that galaxies evolve with time. 
We  model the number density of galaxies as a function of the velocity dispersion $\sigma_v$, taking into account the evolution of galaxies with redshift.  For this, we use the results of Torrey et al. \cite{2015MNRAS.454.2770T} based on the Illustris hydrodynamical simulation. We  use the values in Table 6 in the ArXiv version of 
Ref.~\cite{2015MNRAS.454.2770T}, but we do not use the fitting formula given there. The correct fit is~\cite{Paul} (see also \cite{Cusin:2019eyv})
\begin{align}\label{fit2}
&\log_{10} N(>\sigma_v, z)=\\
&=A(z)+\alpha(z) 
\left(\log_{10}\sigma_v-\gamma(z) \right)+\beta(z)\left(\log_{10}\sigma_v-\gamma( z)\right)^2 -\left(\sigma_v\times10^{-\gamma(z)}\right)^{1/\ln(10)}\,,\nonumber
\end{align}
where the numerical value of $\sigma_v$ is to be taken in units of $\mathrm{km\, s^{-1}}$ and $N$ is in units of $\mathrm{Mpc}^{-3}$. We stress that here $\sigma_v$ represents the \emph{redshift independent} velocity dispersion.  The functions $A$, $\al$, $\beta$ and $\gamma$ are modeled as
\begin{align}
A&=a_0+a_1 z+a_2 z^2\,,\\
\alpha&=\alpha_0+\alpha_1 z+\alpha_2 z^2\,,\\
\beta&=\beta_0+\beta_1 z+\beta_2 z^2\,,\\
\gamma&=\gamma_0+\gamma_1 z+\gamma_2 z^2\,.
\end{align}
For completeness, in the following table we list the coefficients of the fitting formula: 
\begin{table}[ht!]
\centering
\begin{tabular}{c|ccc}
$i$& $i=0$ &  $i=1$&  $i=2$  \\
 \hline
 $A_i$& 7.391498 &5.729400  &  -1.120552  \\
 $\alpha_i$ &-6.863393  &  -5.273271&  1.104114  \\
$\beta_i$ &2.852083  &1.255696  &  -0.286638  \\
 $\gamma_i$ &  0.067032& -0.048683 &  0.007648\\
 \hline
\end{tabular}
\end{table}
 
\begin{figure}[ht!]
        \begin{center}
        \includegraphics[width=0.48\textwidth]{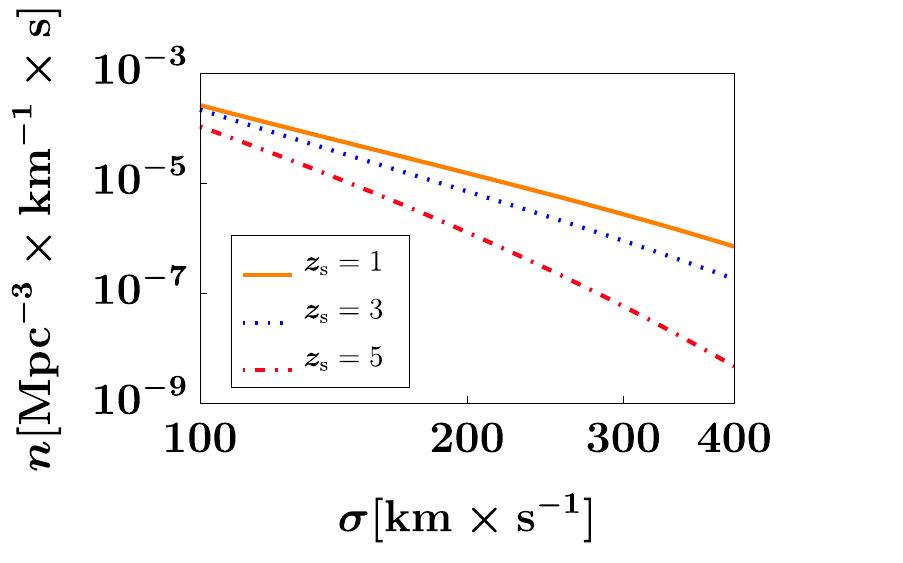}  \includegraphics[width=0.48\textwidth]{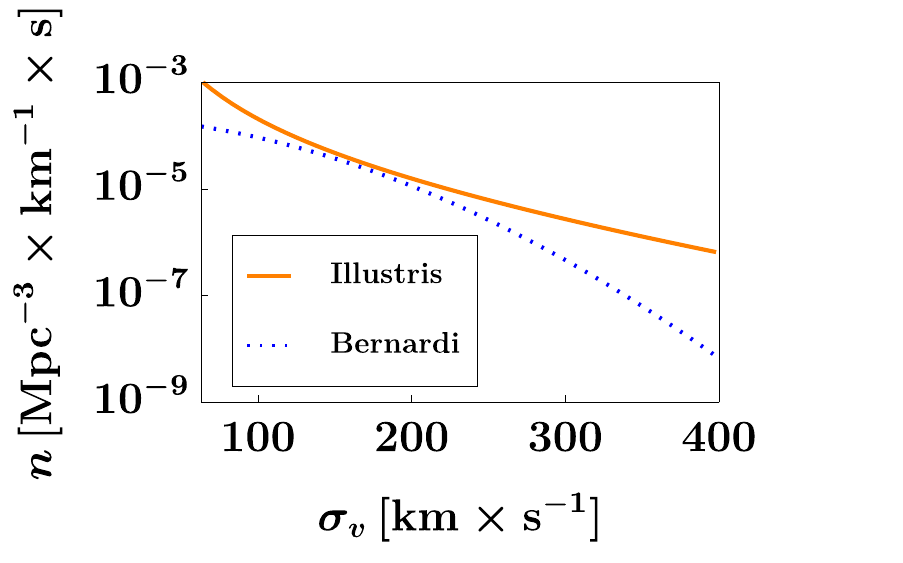}
        \end{center}
        \caption{\label{NN} Left: The comoving number density of galaxies as a function of  of velocity dispersion  $\si_v$ for different redshifts. The figure is obtained from fits to the Illustris simulation  of \cite{2015MNRAS.454.2770T}. Right: Comparison between number density at $z=0$ of \cite{Bernardi_2010} and of \cite{2015MNRAS.454.2770T}. Notice that the latter distribution is valid only for $\sigma_v>10^{1.8}$ km/s. {We notice that the (simulated) catalogue of Illustris \cite{2015MNRAS.454.2770T} contains much more galaxies at $z=0$ than the (observed) sample of \cite{Bernardi_2010} and it shows a different $\sigma_v$ dependence (the latter highly depends on the sample considered).} }
    \end{figure}

The central stellar velocity dispersion,  $\sigma_v$ is defined as the three-dimensional standard deviation of the stellar velocity within the stellar half-mass radius. We   define the comoving number density of galaxies as 
 \be
n^{\text{com}}(\sigma_v, z)\equiv - \frac{\dd N}{\dd\sigma_v}(>\sigma_v, z)\,.
\ee
The result for different redshifts is shown in Fig. \ref{NN}. 

\subsection{Cosmological parameters and definition of distances}
In this appendix, we give the values of cosmological parameters used to derive numerical predictions. 
We assume a $\Lambda$CDM universe with Hubble function
\begin{equation}
H(z) = H_0 \sqrt{\Omega_m (1+z)^3 + \Omega_r (1+z)^4 + \Omega_\Lambda}\,, 
\end{equation}
where 
\begin{align*}
\Omega_{\textrm{m}} &= 0.274\, \\
\Omega_{\textrm{r}} &= 2.5\times 10^{-5} h^{-2}\, \\
\Omega_\Lambda &= 1- \Omega_{\textrm{m}} -\Omega_{\textrm{r}}\, \textrm{and}\\
H_0 &= 100h\;  \mathrm{km \times s^{-1} \times Mpc^{-1}} \,, \\
h&=0.705\,.
\end{align*}
Finally, we recall how the angular distances are defined as functions of comoving distances $r(z_1, z_2)$, 
\begin{equation}
\DL = \frac{r(0,z)}{1+z}\,, \quad
\DS = \frac{r(0,\zS)}{1+\zS}\,, \quad
\DLS = \frac{r(z,\zS)}{1+\zS}\,,\quad
r(z_1,z_2) = \int_{z_1}^{z_2} \frac{1}{H(z)}\;  \mathrm{d}z\,.
\end{equation}
Here $z$ is the redshift of the lens and $\zS$ is the redshift of the source.

\newpage
\bibliography{refs}
\bibliographystyle{utcaps}

\end{document}